%% file: main.tex
\pdfoutput=1
\newif\ifextended\extendedtrue
\documentclass[twocolumn,amsthm]{autart}

\usepackage{amsmath,amssymb}
\usepackage{graphicx}
\usepackage{booktabs}
\usepackage{natbib}%

\makeatletter
\newcommand{\doi}[1]{\@gobble}

\makeatother
\usepackage{url}
\newif\ifxrlinks \xrlinksfalse
\ifextended
 \usepackage{tikz}
 \newcommand{\appref}[1]{Appendix~\ref{#1}}
 \newcommand{\apptabref}[1]{Table~\ref{#1}}
\else
 \ifxrlinks
  \usepackage{xr}
  \newcommand{\appref}[1]{Appendix~\ref{X-#1} of~EV}
  \newcommand{\apptabref}[1]{Table~\ref{X-#1} of~EV}
 \else
  \newcommand{\EVnum}[1]{\ifcsname EV@#1\endcsname
   \csname EV@#1\endcsname\else\textbf{??}\fi}
  \newcommand{\appref}[1]{Appendix~\EVnum{#1} of~EV}
  \newcommand{\apptabref}[1]{Table~\EVnum{#1} of~EV}
  \expandafter\def\csname EV@app:certificate\endcsname{A}
  \expandafter\def\csname EV@tab:certificate-matrices\endcsname{A.1}
  \expandafter\def\csname EV@supp:lower-bound\endcsname{B.1}
  \expandafter\def\csname EV@supp:wellposed\endcsname{B.2}
  \expandafter\def\csname EV@supp:capture\endcsname{B.4}
  \expandafter\def\csname EV@supp:accuracy\endcsname{B.5}
  \expandafter\def\csname EV@supp:simulations\endcsname{B.6}
 \fi
\fi
\defcitealias{ZapfExtended}{EV}

\newcommand{\spow}[2]{\lfloor #1\rceil^{#2}}
\DeclareMathOperator{\sgn}{sgn}
\newcommand{\phibar}{\bar\phi}
\newcommand{\R}{\mathbb R}
\newcommand{\dd}{\,\mathrm d}

\begin{document}

\begin{frontmatter}
\runtitle{Implicit discretization of PID-like sliding-mode control}

\title{Implicit discretization of PID-like sliding-mode control with third-order output accuracy\thanksref{footnoteinfo}}
\ifextended
\thanks[footnoteinfo]{Extended version of a paper submitted to Automatica. Appendices~\ref{app:certificate} and~\ref{supp:proofs} are not part of the journal version. This research did not receive any specific grant from funding agencies in the public, commercial, or not-for-profit sectors. Corresponding author M.~Zapf.}
\else
\thanks[footnoteinfo]{This paper was not presented at any IFAC meeting. The implicit step of Section~\ref{sec:controller} and the implementation of Section~\ref{subsec:cdc} appeared in \citet{Zapf2025}. This research did not receive any specific grant from funding agencies in the public, commercial, or not-for-profit sectors. Corresponding author M.~Zapf.}
\fi

\author[SHS,TUG]{Martin Zapf}\ead{martin.zapf@siemens-healthineers.com},
\author[TUG]{Benedikt Andritsch},
\author[UNAM]{Leonid Fridman},
\author[TUG]{Martin Horn}

\address[SHS]{Siemens Healthineers AG, Kemnath, Germany}
\address[TUG]{Institute of Automation and Control, Graz University of
Technology, Graz, Austria}
\address[UNAM]{Facultad de Ingenier\'ia, Universidad Nacional Aut\'onoma de
M\'exico, Mexico City, Mexico}


\begin{abstract}
A digital PID-like sliding-mode controller is proposed for a  double integrator with a Lipschitz matched disturbance. A proper implicit discretization, matched to the exact sampled dynamics of the plant, attains the accuracy of the continuous-time controller without discretization chattering. Global finite-step convergence to a discrete-time sliding mode is established for two gain designs and every sampling time, using an implicit Lyapunov function and a computer-assisted certificate in exact arithmetic. The resulting controller provides third-order output accuracy, second-order accuracy for its derivative, and exact finite-time rejection of constant matched disturbances. These accuracy orders are proved to be optimal among deterministic causal controllers using the same sampled information.
\end{abstract}

\end{frontmatter}

\section{Introduction}
\label{sec:introduction}
Conventional sliding-mode control achieves robustness against matched
disturbances through discontinuous control action, which generates
chattering in the presence of parasitic dynamics and excites unmodeled
actuator dynamics. The development of continuous higher-order
sliding-mode controllers was motivated by the possibility of  eliminating the discontinuity of the control
signal while simultaneously increasing the sliding order and the
corresponding accuracy~\citep{Fridman2015}.

In particular, the continuous terminal sliding-mode controller proposed
by \citet{Kamal2016} and the continuous twisting algorithm developed by
\citet{TorresGonzalez2017} demonstrated that third-order sliding modes
can be achieved for uncertain second-order systems using only direct
measurements of the output and its first derivative. Thus, no additional
measurements are required compared with conventional sliding-mode
control of second-order systems. The extended closed-loop dynamics of
these algorithms are homogeneous of degree $-1$ with weights $(3,2,1)$,
corresponding to third-order sliding accuracy with respect to the
output~\citep{Levant2005}.

Although these pioneering results established the existence of
continuous third-order sliding-mode controllers, these algorithms
were relatively complicated, and the sufficient stability conditions led
to highly conservative controller gains. This made their practical
tuning difficult and motivated the development of a simpler structure,
familiar to practicing engineers. \citet{PerezVentura2021} introduced a
continuous PID-like sliding-mode controller with only three gains,
retaining the same third-order sliding properties. Using
describing-function and harmonic-balance analysis, two gain designs were
obtained by minimizing, respectively, the amplitude of actuator-induced
chattering and the average power associated with the resulting
oscillations. The stability of both designs was established using
homogeneous Lyapunov functions.

The next natural challenge is the digital implementation of the PID-like
controller while preserving the third-order sliding accuracy associated
with its homogeneity weights $(3,2,1)$ and without reintroducing
chattering. However, straightforward Euler
discretization of higher-order homogeneous sliding-mode algorithms does
not generally preserve their properties:
for controllers, it keeps the error orders with respect to the
sampling time $T$~\citep{Levant2005,Levant2011} but usually causes discretization
chattering~\citep{Andritsch2026}, and its solutions need not
converge to zero~\citep{Efimov2017}.

Such difficulties motivated the development of proper discretization
methods, which preserve the accuracy of the continuous-time algorithm,
for homogeneous differentiators by \citet{LivneLevant2014} and
\citet{Barbot2020}, as well as consistent discretization methods for
homogeneous controllers by \citet{Polyakov2023}. In parallel,
chattering-free implicit implementations of sliding-mode controllers
were investigated by \citet{AcaryBrogliato2010} and
\citet{HuberTwisting2016}. For relative degree one, both lines meet in
the proper implicit discretization of the super-twisting controller by
\citet{SeeberAndritsch2025}, which attains the best possible worst-case
accuracy. These results demonstrate that digital
implementation requires special care to preserve the desired properties
of the continuous-time algorithms.

For relative degree two, implicit Euler has been applied to
the continuous terminal algorithm~\citep{Xiong2019}, implementations
free of discretization chattering exist for arbitrary
order~\citep{AndritschCDC2024}, and the PID-like controller has an
implicit implementation~\citep{Zapf2025}. The latter avoids
discretization chattering in simulations but is not proper: under a
constant disturbance $d$, its only equilibrium combines the correct input $u=-d$
with the output offset $x_1=T^2d$, and no solution does better
(Section~\ref{subsec:cdc}). The offset is of order $T^2$ and grows with
the disturbance amplitude.

This paper proposes a new proper implicit discretization scheme
for the homogeneous PID-like sliding-mode controller, specifically
designed to preserve third-order sliding accuracy in the sampled-data
closed loop. Unlike a straightforward discretization of the
continuous-time equations, the proposed scheme accounts for the exact
zero-order-hold dynamics of the second-order plant through a modified
output, extrapolated by half a sampling
time, and a causal reconstruction of the disturbance
average~\citep{Su2000}; it extends the scheme
of~\citet{SeeberAndritsch2025} to relative degree two. For two certified
gain designs and their homogeneous scaling families, one of them a
design of \citet{PerezVentura2021}, global finite-step convergence to a
discrete-time sliding mode is established for every sampling time
using an implicit Lyapunov function~\citep{Polyakov2015} and a
computer-assisted stability certificate. In
this sliding mode and from the fourth sample on, the physical closed
loop achieves the accuracy orders $O(T^3)$ for the output, $O(T^2)$ for
its first derivative, and $O(T)$ for the disturbance-compensation error,
with explicit sampled and intersample bounds. A causal lower-bound
result further proves that the first two accuracy orders are optimal
among deterministic causal controllers using the same sampled
information. Constant matched disturbances are rejected exactly after a
finite transient, and the closed loop is free of discretization
chattering (Section~\ref{subsec:deadbeat}).

The remainder of the paper is organized as follows.
Section~\ref{sec:problem} presents the sampled-data model, the
accuracy limit of causal controllers and the motivation for the
proposed discretization. Section~\ref{sec:proper} introduces the
proper implicit PID-like discretization scheme and states the main results.
Section~\ref{sec:capture} develops the global finite-step convergence
analysis, while Section~\ref{sec:accuracy} establishes the
sampled-data accuracy orders and corresponding intersample bounds.
Section~\ref{sec:simulations} presents numerical comparisons, Section~\ref{subsec:sim-limits}
examines model errors and measurement noise, and
Section~\ref{sec:conclusions} concludes the paper. \ifextended The remaining proofs\else Complete technical proofs\fi, the computer-assisted Lyapunov certificates, and their
verification procedures are collected in \ifextended Appendices~\ref{app:certificate}
and~\ref{supp:proofs}\else the extended
version~\citep{ZapfExtended}, cited as EV\fi.

\emph{Notation.} We use $\spow{z}{p}=|z|^p\sgn(z)$ for $p>0$, with
$\sgn(0)=0$, and the set-valued sign $\spow{z}{0}=\{\sgn(z)\}$ for $z\ne0$,
while $\spow{0}{0}=[-1,1]$.\ifextended\ The norm $\|\cdot\|$ is Euclidean and, for
matrices, its induced norm.\fi

\section{Motivation and problem statement}
\label{sec:problem}
Consider the double integrator
\begin{equation}
 \dot x_1(t)=x_2(t),\quad \dot x_2(t)=u(t)+\phi(t),\quad t\ge0,
 \label{eq:plant}
\end{equation}
where $x_1(t)\in\R$ is the regulated output, $x_2(t)\in\R$ its
derivative, $u(t)\in\R$ the control input and $\phi(t)\in\R$ the matched
disturbance. The disturbance is unknown but has bounded slope,
\begin{equation}
 |\phi(t)-\phi(t')|\le\Delta|t-t'|\qquad(t,t'\ge0),
 \label{eq:disturbance-class}
\end{equation}
with a constant $\Delta\ge0$, so $|\dot\phi|\le\Delta$ almost everywhere,
and no amplitude bound; only the gain choice uses $\Delta$
(Section~\ref{sec:results}).
Sampling is uniform, at $t_k=kT$, with a constant and known sampling time $T>0$, and
the samples of the plant are $x_{i,k}=x_i(t_k)$. Both states are measured without noise, the input is
unrestricted, and the initial state is arbitrary. The controller
computes $u_k$ from the measurements up to $t_k$ without delay, and the
input is applied through a zero-order hold,
\begin{equation}
 u(t)=u_k\quad\text{on }[t_k,t_{k+1}),\qquad k\ge0.
 \label{eq:hold}
\end{equation}
Integrating \eqref{eq:plant} twice over one sampling interval and
exchanging the order of integration in the double integral of $\phi$
gives the exact sampled model of the plant under
\eqref{eq:hold}. With the interval average
\begin{equation}
 \phibar_k=T^{-1}\int_0^T\phi(t_k+\tau)\dd\tau,\qquad k\ge0,
 \label{eq:average}
\end{equation}
it reads
\begin{align}
 x_{1,k+1}&=x_{1,k}+Tx_{2,k}+\tfrac12T^2u_k\nonumber\\
       &\quad+\int_0^T(T-\tau)\phi(t_k+\tau)\dd\tau,\label{eq:zoh-x1}\\
 x_{2,k+1}&=x_{2,k}+T(u_k+\phibar_k).
 \label{eq:zoh-x2}
\end{align}
The derivative update \eqref{eq:zoh-x2} depends on the disturbance
only through its average, whereas the output update \eqref{eq:zoh-x1}
depends on the convolution of $\phi$ with the ramp $T-\tau$, which the
average does not determine.

A closed loop has the \emph{accuracy orders} $(3,2)$ with respect to $T$
if $|x_{1,k}|\le C_1\Delta T^3$ and $|x_{2,k}|\le C_2\Delta T^2$
after a finite number of steps, with constants $C_1$, $C_2$ independent
of $T$, of the initial state and of the disturbance, which enters only
through $\Delta$~\citep[cf.][]{Levant2005}. The objective is these orders
on the plant, with bounds between the samples and exact
rejection of constant disturbances for all $t$ after a finite time.

\subsection{Best possible worst-case accuracy}
The next proposition, the double-integrator counterpart of
\citet[Proposition~2.1]{SeeberAndritsch2025}, bounds the best possible
worst-case sampled errors of deterministic causal controllers from below.
As there and in \citet{AndritschKochMoreno2026}, its proof compares two
mirrored disturbances that the controller cannot distinguish.
\begin{prop}[Lower bound for causal controllers]
\label{prop:order-lower-bound}
Consider the plant \eqref{eq:plant} under the zero-order hold \eqref{eq:hold} and
any deterministic causal controller, in which $u_k$ is a function of
the samples $x_{1,j},x_{2,j}$ with $j\le k$ and of the inputs $u_j$ with
$j<k$. For every $T>0$, $\Delta>0$ and initial plant state, there is a
disturbance satisfying \eqref{eq:disturbance-class} with
$\limsup_{k\to\infty}|x_{1,k}|\ge7\Delta T^3/24$, and there is one with
$\limsup_{k\to\infty}|x_{2,k}|\ge3\Delta T^2/4$.
\end{prop}
\begin{proof}
Fix the controller and the initial state; the disturbance is
constructed against them. By \eqref{eq:zoh-x1}--\eqref{eq:zoh-x2} and
causality, two disturbances whose difference has zero average and zero
convolution with $T-\tau$ on every $[t_j,t_{j+1}]$ with $j<K$ give the
same samples up to index $K$ and the same inputs up to $u_K$. One such
difference is $2\psi$, with $\psi$ the $T$-periodic function given by
$\psi(t)=\Delta(|t-T/2|-T/4)$ on $[0,T]$: it has slopes $\pm\Delta$, zero
average and midpoint symmetry on each interval, hence zero convolution,
and $\psi(t_k)=\Delta T/4$. For a constant $d$ and an index $K$, let the
branches be $\phi_\pm=d\pm\psi$ on $[0,t_K]$, the hidden phase, and
$\phi_\pm(t)=d\pm(\Delta T/4+\Delta(t-t_K))$ on $[t_K,t_{K+1}]$, the
reveal, continued as constants. Both satisfy \eqref{eq:disturbance-class} and
give the same samples and inputs up to $u_K$, whereas by
\eqref{eq:zoh-x1} their samples $x^\pm_{1,K+1}$ differ by
${2\int_0^T(T-\tau)(\Delta T/4+\Delta\tau)\dd\tau}=7\Delta T^3/12$,
so one of them has $|x_{1,K+1}|\ge7\Delta T^3/24$. Likewise,
\eqref{eq:zoh-x2} gives the difference $\frac32\Delta T^2$ in $x_{2,K+1}$,
so one of them has $|x_{2,K+1}|\ge3\Delta T^2/4$. For the $\limsup$,
one disturbance repeats hidden phases and reveals and keeps at each
reveal the branch with the larger error. The
branches coincide before each hidden phase, and their
difference approaches $2\psi$ over a growing number of intervals, so
each reveal starts arbitrarily close to $\Delta T/4$
(\appref{supp:lower-bound}).
\end{proof}
A guarantee
$|x_{1,k}|\le C\Delta T^p$ for all $k\ge k_0(\phi)$, all disturbances $\phi$
satisfying \eqref{eq:disturbance-class} and all $T\in(0,T_0)$, with $C$
independent of $T$, would give $7\Delta T^3/24\le C\Delta T^p$, which
fails for $p>3$ as $T\to0$; likewise $p>2$ is excluded for $|x_{2,k}|$.
The orders $(3,2)$ of Theorem~\ref{thm:accuracy} are therefore optimal;
another controller can improve only its constants $C_1=7/8$ and
$C_2=23/12$, down to $7/24$ and $3/4$ at best.

\subsection{The PID-like controller and its implicit step}
\label{sec:controller}
The continuous-time law of \citet{PerezVentura2021},
\begin{equation}
 \begin{gathered}
 u=-k_1\spow{x_1}{1/3}-k_2\spow{x_2}{1/2}+v,\\
 \dot v\in-k_3\spow{x_1}{0},
 \end{gathered}
 \label{eq:continuous-pid}
\end{equation}
with positive gains, provides the structure of the digital controller.
Substitution into \eqref{eq:plant} shows that $v$ and $\phi$ enter
$\dot x_2$ only through the sum $w=v+\phi$, so the closed loop is
\begin{equation}
 \begin{gathered}
 \dot x_1=x_2,\qquad
 \dot x_2=-k_1\spow{x_1}{1/3}-k_2\spow{x_2}{1/2}+w,\\
 \dot w\in-k_3\spow{x_1}{0}+\dot\phi.
 \end{gathered}
 \label{eq:continuous-closed-loop}
\end{equation}
Solutions of \eqref{eq:continuous-closed-loop} are understood in the
sense of \citet{Filippov1988}. The third-order sliding set of \eqref{eq:continuous-closed-loop}, on which $x_1$, $x_2$ and $\dot x_2$
vanish, is $x_1=x_2=w=0$: there the integral state cancels the whole
disturbance.

\label{subsec:common-step}
For real $a$, $b$, $w$ in the roles of the output, its derivative
and the integral state plus disturbance, let
$(X,Y,\sigma)=\Phi_T(a,b,w)$ denote the solution of the inclusion
\begin{subequations}
\label{eq:generic-step}
\begin{align}
 X&=a+TY,\qquad \sigma\in\spow{X}{0},\label{eq:step-first}\\
 Y&=b+T[-k_1\spow{X}{1/3}-k_2\spow{Y}{1/2}+w-Tk_3\sigma],\label{eq:step-second}
\end{align}
\end{subequations}
unique by \citet[Theorem~1]{Zapf2025} and by
Proposition~\ref{prop:wellposed}, proved in \appref{supp:wellposed}.
It is the implicit Euler step of length $T$ of
\eqref{eq:continuous-closed-loop} with $\dot\phi=0$ and the successor
$w-Tk_3\sigma$ of $w$ substituted \citep{Zapf2025}: every term is
evaluated at the successor, so $X$ and $Y$ appear on both sides. Both
controllers below use it with $v_{k+1}=v_k-Tk_3\sigma_k$. Where the successor output is zero, the
step can select any $\sigma\in[-1,1]$, whereas an explicit step switches
between $\pm1$; this is how implicit discretization avoids discretization
chattering~\citep{AcaryBrogliato2010}.

\begin{prop}[Explicit form of the implicit step]
\label{prop:wellposed}
For every $T,k_1,k_2,k_3>0$ and $(a,b,w)\in\R^3$,
\eqref{eq:generic-step} has a unique solution. Define the selection
variable
\begin{equation}
 s=\frac{a/T+b+k_2\sqrt T\,\spow{a}{1/2}+Tw}{T^2k_3}.
 \label{eq:selection}
\end{equation}
For $|s|\le1$, the \emph{sliding branch}, this solution is
$(X,Y,\sigma)=(0,-a/T,s)$.
For $|s|>1$, the solution has $\sigma=\sgn(s)$ and $X=a+TY$, where $Y$ is the
unique root of $R(\cdot\,;\sigma)$,
\begin{equation}
 \begin{split}
 R(y;\sigma)={}&y+Tk_1\spow{a+Ty}{1/3}+Tk_2\spow{y}{1/2}\\
                &-b-Tw+T^2k_3\sigma,
 \end{split}
 \label{eq:scalar-residual}
\end{equation}
and $R$ is continuous and strictly increasing in $y$.
\end{prop}

\subsection{Earlier implementation and its offset equilibrium}
\label{subsec:cdc}
The earlier implementation of \citet{Zapf2025}, whose offset motivates
the proposed controller, feeds the measured samples
directly to the solution map $\Phi_T$ of \eqref{eq:generic-step}:
\begin{subequations}
\label{eq:cdc-implementation}
\begin{gather}
 (X_k,Y_k,\sigma_k)=\Phi_T(x_{1,k},x_{2,k},v_k),\nonumber\\
 v_{k+1}=v_k-Tk_3\sigma_k,\label{eq:cdc-step}\\
 u_k=-k_1\spow{X_k}{1/3}-k_2\spow{Y_k}{1/2}+v_{k+1}.\label{eq:cdc-input}
\end{gather}
\end{subequations}
Here $X_k$ and $Y_k$ are predictions of $x_{1,k+1}$ and
$x_{2,k+1}$: \eqref{eq:generic-step} and
\eqref{eq:cdc-input} give $Y_k=x_{2,k}+Tu_k$ and
$X_k=x_{1,k}+TY_k=x_{1,k}+Tx_{2,k}+T^2u_k$, a model that omits the disturbance from
\eqref{eq:zoh-x1}--\eqref{eq:zoh-x2} and carries the input into the
output with coefficient $T^2$ instead of $T^2/2$.

For $\phi=d$ constant, an equilibrium requires $u=-d$ and $x_2=0$ by
\eqref{eq:zoh-x1}--\eqref{eq:zoh-x2}, and a constant $v$ requires
$\sigma=0$ in \eqref{eq:cdc-step}, hence $X=0$ by \eqref{eq:step-first}.
Then $Y=x_2+Tu=-Td$, $x_1=X-TY=T^2d$, and \eqref{eq:cdc-input} gives $v$:
for all positive gains, $T>0$ and $d$, the unique equilibrium is
\begin{equation}
 \begin{aligned}
 &x_1=T^2d,\quad x_2=0,\quad u=-d,\\
 &v=-d-k_2\sgn(d)\sqrt{T|d|}.
 \end{aligned}
 \label{eq:cdc-offset}
\end{equation}
The values $(X,Y,\sigma)=(0,-Td,0)$ and this $v$ satisfy
\eqref{eq:generic-step}, and $u+d=0$ keeps the plant at rest. The offset
is the one-step prediction error, $x_1-X=-T^2u=T^2d$.

No solution does better. By \eqref{eq:zoh-x2},
$Y_k=x_{2,k+1}-Td$ and $X_k=x_{1,k}+Tx_{2,k+1}-T^2d$. If
$|x_{1,k}|+T|x_{2,k+1}|<T^2|d|$ held from some index on, then
$\sigma_k=-\sgn(d)$ there, and $|v_k|$ would grow without bound by
\eqref{eq:cdc-step}. So would $|u_k|$ by \eqref{eq:cdc-input}, as $X_k$ and $Y_k$ stay bounded;
but $u_k=(x_{2,k+1}-x_{2,k})/T-d$ stays bounded, a contradiction. Hence
$\limsup_{k\to\infty}(|x_{1,k}|+T|x_{2,k+1}|)\ge T^2|d|$ for every
solution. For $d\ne0$, where $\Delta=0$, the earlier implementation thus
neither rejects the disturbance exactly nor has accuracy orders in the
sense of Section~\ref{sec:problem}. Section~\ref{sec:simulations} shows
it approaching \eqref{eq:cdc-offset}, whose attractivity is not analyzed.
The third-order accuracy stated in \citet{Zapf2025} for the undisturbed
plant thus does not extend to disturbed plants.

\section{Proper implicit discretization}
\label{sec:proper}

\subsection{The proposed controller}
\label{subsec:measured-combination}
The proposed controller replaces the model of Section~\ref{subsec:cdc} by
a causal prediction whose error depends only on the deviation of the
disturbance from its average over the preceding sampling interval, as
\citet{SeeberAndritsch2025} do for relative degree one.

\emph{The coordinate the plant can follow.} A single held input can in
general make the plant match only one of the two successors $X,Y$ of the
implicit step. For a transformed coordinate
$c_k^\alpha=x_{1,k}+\alpha Tx_{2,k}$, computed from the measurements,
\eqref{eq:zoh-x1}--\eqref{eq:zoh-x2} give, for every held input and
every disturbance,
\begin{equation}
 \begin{split}
 c_{k+1}^\alpha={}&c_k^\alpha+Tx_{2,k+1}+(\alpha-\tfrac12)T^2(u_k+\phibar_k)\\
 &+\int_0^T(\tfrac T2-\tau)\phi(t_k+\tau)\dd\tau.
 \end{split}
 \label{eq:combination-advance}
\end{equation}
Since $\int_0^T(\tfrac T2-\tau)\dd\tau=0$, the integral vanishes for
constant $\phi$, and subtracting $\phi(t_k+T/2)$ inside the integral bounds it by
$\Delta T^3/12$. Only
$\alpha=1/2$ removes the input term, so that the plant realizes
$X=a+TY$ in \eqref{eq:step-first} exactly for constant disturbances,
whatever input is held; for $\alpha=-1/2$, the terms with $x_{2,k+1}$
and $u_k$ add up to $Tx_{2,k}$, and the input cannot influence
$c_{k+1}^\alpha$. The
coordinate is the output extrapolated by $T/2$,
\begin{equation}
 c_k=x_{1,k}+\tfrac T2x_{2,k}.
 \label{eq:measured-combination}
\end{equation}
At $t_k$, the current average $\phibar_k$ in $x_{2,k+1}$ is unknown; the
preceding one follows from \eqref{eq:zoh-x2}:
\begin{equation}
 \phibar_{k-1}=\frac{x_{2,k}-x_{2,k-1}}T-u_{k-1},\qquad k\ge1.
 \label{eq:reconstruct-average}
\end{equation}

\emph{The input that realizes the implicit step.} The proposed
controller propagates three controller states by the implicit step
\eqref{eq:generic-step}: the modified sliding variable $a$, named as
in~\citet{SeeberAndritsch2025}, its difference quotient $b$ and the
integral state $v$, with $a$ and $b$ in the roles of $x_1$ and $x_2$. With the
initialization
\begin{equation}
 a_0=c_0,\quad b_0=x_{2,0},\quad v_0\in\R,\quad \phibar_{-1}=0,
 \label{eq:startup}
\end{equation}
the solution map $\Phi_T$ of \eqref{eq:generic-step} chooses their
successors, and the measurements enter only through the averages:
\begin{subequations}
\label{eq:proposed-controller}
\begin{align}
 (a_{k+1},b_{k+1},\sigma_k)&=\Phi_T(a_k,b_k,v_k+\phibar_{k-1}),\label{eq:proposed-step}\\
 v_{k+1}&=v_k-Tk_3\sigma_k.\label{eq:proposed-v}
\end{align}
\end{subequations}
For these states, \eqref{eq:step-first} and \eqref{eq:step-second} read
\begin{align}
 b_{k+1}&=(a_{k+1}-a_k)/T,
 \label{eq:retained-difference}\\
 b_{k+1}&=b_k+T[-k_1\spow{a_{k+1}}{1/3}
                  -k_2\spow{b_{k+1}}{1/2}\nonumber\\
 &\qquad+v_{k+1}+\phibar_{k-1}].
 \label{eq:modified-second}
\end{align}
The causal prediction of $c_{k+1}$ drops the integral in
\eqref{eq:combination-advance} and replaces $\phibar_k$ by
$\phibar_{k-1}$, so it is exact for $\phi$ constant on
$[t_{k-1},t_{k+1}]$. The held input $u_k$ sets this prediction equal to
$a_{k+1}$:
\begin{gather}
 c_k+Tx_{2,k}+T^2(u_k+\phibar_{k-1})=a_{k+1},
 \label{eq:modified-sliding-variable}\\
 \begin{aligned}
 u_k&=\frac{a_{k+1}-c_k}{T^2}-\frac{x_{2,k}}T-\phibar_{k-1}\\
     &=\frac{a_{k+1}-x_{1,k}}{T^2}-\frac{3x_{2,k}}{2T}-\phibar_{k-1}.
 \end{aligned}
 \label{eq:applied-input}
\end{gather}
With \eqref{eq:reconstruct-average} and \eqref{eq:startup}, equations
\eqref{eq:proposed-controller}--\eqref{eq:applied-input} define the
proposed controller from quantities available at $t_k$. Subtracting
\eqref{eq:modified-sliding-variable} from \eqref{eq:combination-advance}
with $\alpha=1/2$ gives, for $k\ge0$,
\begin{equation}
 \begin{split}
 c_{k+1}-a_{k+1}
 ={}&\int_0^T(\tfrac T2-\tau)\phi(t_k+\tau)\dd\tau\\
 &+T^2(\phibar_k-\phibar_{k-1}),
 \end{split}
 \label{eq:physical-modified-difference}
\end{equation}
which vanishes from $k=1$ on for constant $\phi$.

\ifextended
\emph{Why the extrapolated output is needed.} If the earlier
implementation \eqref{eq:cdc-implementation}, which acts on $x_{1,k}$,
receives $v_k+\phibar_{k-1}$ in place of $v_k$, the average removes
the offset, but on the sliding branch the characteristic polynomial
$z^2-z/2+1/2$ has roots of modulus $1/\sqrt2$: constant disturbances
are rejected at best asymptotically, not after a finite number of steps as
in Theorem~\ref{thm:accuracy}(iii).
\fi

\subsection{Algorithm and PID-like form}
\label{subsec:algorithm}
Algorithm~\ref{alg:controller} is the explicit form of the proposed controller; it
stores only $a_{k+1},b_{k+1},v_{k+1},x_{2,k},u_k$.
\begin{alg}[Proposed controller]
\label{alg:controller}
\upshape
\makeatletter\interlinepenalty=10000 \@beginparpenalty=10000 \makeatother
Initialize $a_0=c_0$, $b_0=x_{2,0}$ and $v_0\in\R$
by~\eqref{eq:startup}. At each sample $t_k$:
\begin{enumerate}
\makeatletter\@itempenalty=10000 \makeatother
\item measure $x_{1,k},x_{2,k}$ and form $c_k$ by
 \eqref{eq:measured-combination};
\item if $k\ge1$, reconstruct $\phibar_{k-1}$ by
 \eqref{eq:reconstruct-average}; else set $\phibar_{-1}=0$;
\item with $(a,b,w)=(a_k,b_k,v_k+\phibar_{k-1})$, compute $s$ by
 \eqref{eq:selection}; if $|s|\le1$, set
 $(a_{k+1},b_{k+1},\sigma_k)=(0,-a_k/T,s)$; otherwise set
 $\sigma_k=\sgn(s)$, let $b_{k+1}$ be the root of
 \eqref{eq:scalar-residual} and $a_{k+1}=a_k+Tb_{k+1}$
 (Proposition~\ref{prop:wellposed});
\item update $v_{k+1}=v_k-Tk_3\sigma_k$ by \eqref{eq:proposed-v};
\item apply and hold $u_k$ from \eqref{eq:applied-input}.
\end{enumerate}
\end{alg}
\emph{PID-like form.} With $a_{k+1}=a_k+Tb_{k+1}$ inserted in
\eqref{eq:applied-input}, the split
$b_{k+1}-x_{2,k}=(b_{k+1}-b_k)+(b_k-x_{2,k})$ and
\eqref{eq:modified-second}, the average terms cancel:
\begin{equation}
 \begin{split}
 u_k={}&-k_1\spow{a_{k+1}}{1/3}-k_2\spow{b_{k+1}}{1/2}+v_{k+1}\\
       &+\frac{a_k-c_k}{T^2}+\frac{b_k-x_{2,k}}T.
 \end{split}
 \label{eq:pid-corrections}
\end{equation}
The input is thus the PID-like law \eqref{eq:continuous-pid} evaluated
at the implicit successor, plus two corrections.
\begin{rem}[Relation to the earlier implementation]
\label{rem:phi-zero}
For $\phi=0$, the plant follows the controller states exactly, $a_k=c_k$
and $b_k=x_{2,k}$, by \eqref{eq:physical-modified-difference},
\eqref{eq:startup}, \eqref{eq:retained-difference} and
\eqref{eq:combination-advance}; the corrections in
\eqref{eq:pid-corrections} vanish, and the proposed controller is the
earlier implementation \eqref{eq:cdc-implementation} on $c_k$
instead of $x_{1,k}$.
\end{rem}

\emph{Capture and accuracy.} The controller states evolve by
\eqref{eq:proposed-controller} alone: the plant acts on them only through
the reconstructed averages. With the sum $q_k=v_k+\phibar_{k-2}$ in
place of $v_k$, their recurrence is the implicit Euler step of the
continuous closed loop \eqref{eq:continuous-closed-loop}, with the
difference quotient of the averages, bounded by $\Delta$, in place of
$\dot\phi$
(Section~\ref{subsec:transformed-recurrence}).
For the certified gain designs below, $(a_k,b_k,q_k)$ reaches zero
after a finite number of steps and stays there, for every $T$
(Theorem~\ref{thm:capture}, proved in Section~\ref{sec:capture}). We call
this \emph{capture}, and the zero state the \emph{discrete-time sliding
mode}. The plant, in turn, follows the controller states: by
\eqref{eq:physical-modified-difference}, $c_k$ follows $a_k$ up to
$O(\Delta T^3)$ from $k=2$ on. Theorem~\ref{thm:accuracy}, the central
result, states the accuracy that results; it is proved in
Section~\ref{sec:accuracy}.

\subsection{Capture theorem and certified designs}
\label{sec:results}
Capture is proved for two gain designs, each a family of gains with a scaling parameter $\Lambda>0$ and a certified disturbance-slope bound (Table~\ref{tab:certified-gains}). Design A with $\Lambda=1$ is the chattering-minimizing set of \citet{PerezVentura2021}, who prove stability of the continuous controller with a homogeneous Lyapunov function computed by sum-of-squares programming. The argument that carries a decrease over to the implicit step, however, needs a Lyapunov function with convex sublevel sets~\citep[Lemma~6.2]{SeeberAndritsch2025}, which is not established for theirs. Each design therefore needs one certificate of decrease (Proposition~\ref{prop:certificate}), the only computer-assisted step of the proof (Section~\ref{subsec:ilf}); a scaling extends it from $\Lambda=1$ to every $\Lambda$ (Section~\ref{subsec:transformed-recurrence}).

Design B lies outside the family of design A and shows that the certification is not specific to one gain set. For the proposed controller, capture from every initial state and for every disturbance satisfying \eqref{eq:disturbance-class} requires $\Delta<k_3$ (Section~\ref{subsec:transformed-recurrence}). Both designs are certified for half of this bound, $\Delta_{\mathrm{cert}}=k_3/2$, so design A covers a slope bound $\Delta$ for $\Lambda\ge20\Delta/11$ and design B for $\Lambda\ge\Delta$; gains outside the two designs need a new certificate of decrease.
\begin{table}[t]
\centering
\caption{Gain designs with the scaling parameter $\Lambda>0$ and their certified disturbance-slope bounds.}
\label{tab:certified-gains}
\begin{tabular}{ccc}
\toprule
Design & $(k_1,k_2,k_3)$ & $\Delta_{\mathrm{cert}}$\\
\midrule
A & $\bigl(2.7\,\Lambda^{2/3},5.345\,\Lambda^{1/2},1.1\,\Lambda\bigr)$ & $0.55\,\Lambda$\\
B & $\bigl(5\Lambda^{2/3},10\Lambda^{1/2},2\Lambda\bigr)$ & $\Lambda$\\
\bottomrule
\end{tabular}
\end{table}

\begin{thm}[Global finite-step capture]
\label{thm:capture}
Let the gains be those of either design of Table~\ref{tab:certified-gains} with some $\Lambda>0$, and let $0\le\Delta\le\Delta_{\mathrm{cert}}$. Consider the plant \eqref{eq:plant} under the hold \eqref{eq:hold}, a disturbance satisfying \eqref{eq:disturbance-class} with this $\Delta$, and the controller \eqref{eq:proposed-controller}--\eqref{eq:applied-input} with reconstruction \eqref{eq:reconstruct-average} and initialization \eqref{eq:startup}. Then, for every $T>0$, every initial plant state and every $v_0\in\R$, there is a least index \mbox{$k_{\mathrm{cap}}\ge2$} such that
\begin{equation}
 a_k=b_k=q_k=0\qquad(k\ge k_{\mathrm{cap}}),
 \label{eq:transformed-capture}
\end{equation}
where $q_k=v_k+\phibar_{k-2}$.
\end{thm}
From $k_{\mathrm{cap}}$ on, $v_k=-\phibar_{k-2}$, the discrete
counterpart of the third-order sliding set: the integral state cancels
the newest average known when $v_k$ is computed. The index is at least $2$
because $\phibar_0$ differs from the initial value
$\phibar_{-1}=0$ by an amount not bounded by $\Delta$;
Theorem~\ref{thm:capture-recurrence} bounds $k_{\mathrm{cap}}$ from above.

\subsection{Sampled and intersample bounds}
\begin{thm}[Sampled and intersample accuracy]
\label{thm:accuracy}
Under the hypotheses of Theorem~\ref{thm:capture}, define
$k_{\mathrm{acc}}=\max\{k_{\mathrm{cap}},3\}$ and the constants
\begin{equation}
 C_1=\frac78,\qquad C_2=\frac{23}{12},\qquad C_3=2C_2+\frac12=\frac{13}{3}.
 \label{eq:accuracy-constants}
\end{equation}
\textup{(i)} For every $k\ge k_{\mathrm{acc}}$, the samples satisfy
\begin{equation}
 |x_{1,k}|\le C_1\Delta T^3,\qquad |x_{2,k}|\le C_2\Delta T^2.
 \label{eq:sampled-bounds}
\end{equation}
The bound on $x_{1,k}$ is attained: for every initial plant state,
$v_0\in\R$, $T>0$ and $0<\Delta\le\Delta_{\mathrm{cert}}$, the ramp
$\phi(t)=\Delta t$ gives $x_{1,k}=C_1\Delta T^3$ for every
$k\ge k_{\mathrm{acc}}$.

\textup{(ii)} For every $t\ge t_{k_{\mathrm{acc}}}$, the states and the input satisfy
\begin{equation}
 \begin{gathered}
 |x_1(t)|\le(C_1+C_3/8)\Delta T^3,\\
 |x_2(t)|\le(C_2+1/8)\Delta T^2,\quad |u(t)+\phi(t)|\le C_3\Delta T.
 \end{gathered}
 \label{eq:intersample-bounds}
\end{equation}

\textup{(iii)} If $\phi(t)=d$ for $t\ge t_*\ge0$, then $x_1(t)=x_2(t)=0$
and $u(t)=-d$ for $t\ge t_{k_*}$, and $v_k=-d$ for $k\ge k_*$, where
$k_*=\max\{k_{\mathrm{acc}},\left\lceil t_*/T\right\rceil+3\}$.
\end{thm}
The constants $C_1$, $C_2$, $C_3$ do not depend on $T$, on the gains, on
the initial data or on the disturbance, which enters the bounds only
through $\Delta$. The index $k_{\mathrm{acc}}$ in general depends on them: for constant
$\phi=d$, capture requires $v=-d$, and $|v_{k+1}-v_k|\le Tk_3$ gives
$t_{k_{\mathrm{cap}}}\ge|v_0+d|/k_3$. The closed
loop thus has the accuracy orders $(3,2)$ of
Section~\ref{sec:problem}, optimal by
Proposition~\ref{prop:order-lower-bound}, and, with
$|u+\phi|\le C_3\Delta T$, the accuracy triple $(3,2,1)$ of a
third-order sliding mode under sampling~\citep{Levant2005}.

\subsection{Deadbeat control and discretization chattering}
\label{subsec:deadbeat}
After capture, $a_{k+1}=0$, and \eqref{eq:applied-input} becomes
$u_k=-x_{1,k}/T^2-3x_{2,k}/(2T)-\phibar_{k-1}$: the two-step deadbeat law
of the exact sampled model \eqref{eq:zoh-x1}--\eqref{eq:zoh-x2} with
one-step-delayed compensation. The loop is then linear, and each sample depends only on the
disturbance over the last three sampling intervals
(Section~\ref{sec:accuracy}).

For the certified designs, the proposed controller is
discretization-chattering-free in the nomenclature of
\citet[Definition~2]{Andritsch2026}: a sampled-data loop with an
asymptotically stable continuous counterpart is so if its origin is
asymptotically stable in the absence of disturbances. What a
disturbance adds is disturbance chattering, which sampling makes
unavoidable. For $\phi=0$, the plant follows the
controller states exactly (Remark~\ref{rem:phi-zero}), so that
$(x_{1,k},x_{2,k},v_k)$ is a linear image of $(a_k,b_k,q_k)$, and these
obey the recurrence of Section~\ref{sec:capture} with zero increments
from $k=0$ on, whose
origin is finite-step stable (Theorem~\ref{thm:capture-recurrence}).
Theorem~\ref{thm:accuracy} bounds the disturbance chattering from
$t_{k_{\mathrm{acc}}}$ on, for the input by $C_3\Delta T$ around the
equivalent control $-\phi$; it vanishes for constant disturbances.

\section{Global finite-step capture}
\label{sec:capture}
This section proves Theorem~\ref{thm:capture} in four steps; the last
three follow \citet[Lemmas~6.2 and~6.5]{SeeberAndritsch2025}. (i)~A
normalization removes $T$ from the recurrence of $(a_k,b_k,q_k)$
(Section~\ref{subsec:transformed-recurrence}). (ii)~A homogeneous
function $V$ with convex sublevel sets, defined implicitly, decreases
along the continuous closed loop; by homogeneity, a computer-assisted
certificate of decrease on one level set suffices (Lemma~\ref{lem:ilf},
Proposition~\ref{prop:certificate}). (iii)~By convexity, $V$ then
decreases along the implicit step for every step size, and every
trajectory enters each sublevel set after finitely many updates
(Lemma~\ref{lem:decrease}). (iv)~In a small sublevel set, the recurrence
is a closed-form map that reaches zero in at most three updates
(Lemma~\ref{lem:local-capture}). Theorem~\ref{thm:capture-recurrence}
combines (ii)--(iv).
\subsection{Normalized recurrence}
\label{subsec:transformed-recurrence}
For $k\ge2$, with the sum $q_k=v_k+\phibar_{k-2}$ of
Theorem~\ref{thm:capture}, define the disturbance increment
\begin{equation}
 \delta_k=(\phibar_{k-1}-\phibar_{k-2})/T.
 \label{eq:proof-coordinates}
\end{equation}
Then $q_{k+1}=v_{k+1}+\phibar_{k-1}$ is the sum that enters
\eqref{eq:modified-second}, and $q_{k+1}-q_k=T(-k_3\sigma_k+\delta_k)$,
so the disturbance acts on $(a_k,b_k,q_k)$ only through $\delta_k$.
Since
$\phibar_{k-1}-\phibar_{k-2}
 =T^{-1}\int_0^T[\phi(t_{k-1}+\tau)-\phi(t_{k-2}+\tau)]\dd\tau$
and the integrand has magnitude at most $\Delta T$ by
\eqref{eq:disturbance-class},
\begin{equation}
 |\delta_k|\le\Delta\qquad(k\ge2).
 \label{eq:increment-bound}
\end{equation}
At $k=1$ the bound can fail: for constant $\phi=d\ne0$, the same
quotient is $d/T$; the analysis therefore starts at $k=2$ from an arbitrary
state.

Equations \eqref{eq:retained-difference}, \eqref{eq:modified-second} and
$q_{k+1}=q_k+T(-k_3\sigma_k+\delta_k)$ are the implicit Euler step of length $T$ of the
closed loop \eqref{eq:continuous-closed-loop},
written for $(a,b,q)$ in place of $(x_1,x_2,w)$ and with
$\delta_k$ in place of $\dot\phi$. The normalization with the homogeneity
weights $(3,2,1)$ of the continuous closed loop,
\begin{equation}
 \zeta_k=(\xi_k,y_k,r_k)^\top=(a_k/T^3,b_k/T^2,q_k/T)^\top,
 \label{eq:normalization}
\end{equation}
removes $T$. Substituting it into the three equations, using
$\spow{T^3\xi}{1/3}=T\spow{\xi}{1/3}$,
$\spow{T^2y}{1/2}=T\spow{y}{1/2}$ and $\spow{T^3\xi}{0}=\spow{\xi}{0}$, and
dividing out the common powers of $T$ gives the unit-step recurrence
\begin{equation}
 \begin{gathered}
 \zeta_{k+1}=\zeta_k+f(\zeta_{k+1},\sigma_k,\delta_k),\\
 \sigma_k\in\spow{\xi_{k+1}}{0},\qquad k\ge2,
 \end{gathered}
 \label{eq:normalized-recurrence}
\end{equation}
with the closed-loop vector field
\begin{equation}
 f(\zeta,\sigma,\delta)
 =\begin{pmatrix}y\\-k_1\spow{\xi}{1/3}-k_2\spow{y}{1/2}+r\\-k_3\sigma+\delta\end{pmatrix}.
 \label{eq:vector-field}
\end{equation}
The recurrence contains no $T$; the sampling time enters only through
$\zeta_2$ and the increments. Indeed, \eqref{eq:normalization}
is $\zeta=D(T)^{-1}(a,b,q)^\top$ with the dilation
$D(\gamma)=\operatorname{diag}(\gamma^3,\gamma^2,\gamma)$, $\gamma>0$, and
$f$ is homogeneous of degree~$-1$,
\begin{equation}
 f(D(\gamma)\zeta,\sigma,\delta)=\gamma^{-1}D(\gamma)f(\zeta,\sigma,\delta),
 \label{eq:weighted-homogeneity}
\end{equation}
with unchanged admissible sign selections. One analysis of
\eqref{eq:normalized-recurrence} thus covers every $T$. It also covers
every $\Lambda$: the field $f_\Lambda$ of the gains of
Table~\ref{tab:certified-gains} satisfies
$f_\Lambda(\Lambda\zeta,\sigma,\delta)=\Lambda f_1(\zeta,\sigma,\delta/\Lambda)$,
so $\zeta_k/\Lambda$ obeys \eqref{eq:normalized-recurrence} for
$\Lambda=1$ with the increments $\delta_k/\Lambda$, which satisfy the
slope bound of $\Lambda=1$\ifextended\ (\appref{supp:remarks})\fi. The rest of this section takes $\Lambda=1$.

The third component of \eqref{eq:normalized-recurrence} is
$r^+=r-k_3\sigma+\delta$ for the successor $\zeta^+=(\xi^+,y^+,r^+)$ of
$\zeta=(\xi,y,r)$; substituting it into the others gives the generic
step $(\xi^+,y^+,\sigma)=\Phi_1(\xi,y,r+\delta)$, so
Proposition~\ref{prop:wellposed} yields a unique successor.

The bound $\Delta<k_3$ of Section~\ref{sec:results} is necessary for the proposed controller.
By \eqref{eq:normalized-recurrence}, a zero state has a zero successor only with $\sigma_k=\delta_k/k_3\in\spow{0}{0}$, and the ramp $\phi(t)=\Delta t$ gives $\delta_k=\Delta$ for all $k\ge2$. For $\Delta>k_3$, the zero state then has no zero successor, and capture cannot persist. For $\Delta=k_3$, initial data reach the fixed points $(\xi,0,k_1\xi^{1/3})$, $\xi>0$, of \eqref{eq:normalized-recurrence} with $\sigma_k=1$, which are never captured, since a step with $\xi_{k+1}\ne0$ has an explicit predecessor.

\subsection{Implicit Lyapunov function and certificate of \mbox{decrease}}
\label{subsec:ilf}
The Lyapunov function and the certificate of decrease serve only the analysis; the controller evaluates neither. The Lyapunov function $V(\zeta)$ is the factor $\gamma$ for which $D(\gamma)^{-1}\zeta$, with the dilation $D(\gamma)$ of Section~\ref{subsec:transformed-recurrence}, lies on the unit level set of a quartic $S$ with convex sublevel sets. For symmetric positive-definite matrices $P_1,P_2,P_3$ and $\chi\in\R^3$, define
\begin{equation}
 \begin{aligned}
 S(\chi)&=\sum_{i=1}^3(\chi^\top P_i\chi)^2,\qquad \mathcal K=\{S\le1\},\\
 J(\chi)&=\nabla S(\chi)^\top W\chi,\qquad W=\operatorname{diag}(3,2,1).
 \end{aligned}
 \label{eq:quartic}
\end{equation}
The function $J$ is the derivative of $S$ along the dilation, $J(\chi)=\frac{\dd}{\dd\gamma}S(D(\gamma)\chi)$ at $\gamma=1$.
\begin{lem}[Convexity]
\label{cert:geometry}
The quartic $S$ is smooth, convex and coercive, with $S(\chi')\ge S(\chi)+\nabla S(\chi)^\top(\chi'-\chi)$ for all $\chi,\chi'\in\R^3$; $\mathcal K$ is compact and convex with zero in its interior.
\end{lem}

Following \citet{Polyakov2015}, the next lemma defines $V$ implicitly. It requires $J>0$ on $\{S=1\}$, which makes every curve $\gamma\mapsto D(\gamma)\chi$ with $\chi\ne0$ cross $\{S=1\}$ exactly once; convexity alone does not ensure this. Lemmas~\ref{cert:geometry} and~\ref{lem:ilf} are proved in \appref{supp:capture}.
\begin{lem}[Implicit Lyapunov function]
\label{lem:ilf}
Let $J\ge m_J>0$ on $\{S=1\}$. For every $\zeta\ne0$, the equation
\begin{equation}
 Q(\zeta,\gamma):=S(D(\gamma)^{-1}\zeta)-1=0
 \label{eq:ilf}
\end{equation}
has a unique positive root $V(\zeta)$. With $V(0)=0$, this function is continuous, positive definite and smooth off zero, satisfies $V(D(\mu)\zeta)=\mu V(\zeta)$ for $\mu>0$, and has compact convex sublevel sets
\begin{equation}
 \{V\le\gamma\}=D(\gamma)\mathcal K\qquad(\gamma>0).
 \label{eq:sublevels}
\end{equation}
These contain a neighborhood of zero and shrink to zero as $\gamma\downarrow0$.
\end{lem}
\begin{prop}[Certified boundary inequalities]
\label{prop:certificate}
For each design, there are rational symmetric positive-definite matrices $P_1,P_2,P_3$ such that, with $\epsilon=1/100$ and $m_J=1/2$ for both designs and whenever $S(\chi)=1$, $\sigma\in\spow{\chi_1}{0}$ and $|\delta|\le\Delta_{\mathrm{cert}}$,
\begin{equation}
 \nabla S(\chi)^\top f(\chi,\sigma,\delta)\le-\epsilon,
 \qquad J(\chi)\ge m_J.
 \label{eq:certificate}
\end{equation}
\end{prop}
The first inequality makes $f$ point into $\mathcal K$ on its boundary $\{S=1\}=\{V=1\}$. Since the dilation leaves $\delta$ unchanged in \eqref{eq:weighted-homogeneity}, this one inequality covers every level set of $V$; Lemma~\ref{lem:decrease} carries it over to the unit-step recurrence \eqref{eq:normalized-recurrence}. The second inequality is the hypothesis of Lemma~\ref{lem:ilf}. The matrices come from a floating-point search and are rounded to rationals; the proof for the rounded matrices is computer-assisted (\appref{app:certificate}). Affinity in $(\sigma,\delta)$ and a symmetry reduce both inequalities to $\sigma=1$ and $\delta=\pm\Delta_{\mathrm{cert}}$ on half of $\{S=1\}$, which twelve rectangular patches parameterize. On these patches, exact rational interval arithmetic bounds both quantities on rectangles, which are bisected until each meets the margins $\epsilon$ and $m_J$; the runs for the two designs end with $45\,209$ and $46\,289$ accepted rectangles. The matrices (\apptabref{tab:certificate-matrices}), the verification program, which needs only the Python standard library, its run records and the simulation scripts are available at \url{https://github.com/MartinZapf/Implicit-PID-like}.

Unlike \citet[Lemma~6.2]{SeeberAndritsch2025}, the next lemma, step~(iii), gives the decrement $\omega$ in closed form, which the bound on $k_{\mathrm{cap}}$ and Corollary~\ref{cor:small-period} use.
\begin{lem}[Decrease and finite entry]
\label{lem:decrease}\label{lem:finite-entry}
For either design, let $J_{\max}\ge m_J$ be the maximum of $J$ on the compact set $\{S=1\}$ and $\omega(\lambda)=\epsilon\lambda/(J_{\max}(\lambda+\epsilon))$. Every update of \eqref{eq:normalized-recurrence} with $|\delta_k|\le\Delta_{\mathrm{cert}}$ satisfies
\begin{equation}
 V(\zeta_k)-V(\zeta_{k+1})\ge\omega(V(\zeta_{k+1})),
 \label{eq:uniform-decrement}
\end{equation}
and the origin is invariant. Hence, for each $\ell>0$, every trajectory with $|\delta_k|\le\Delta_{\mathrm{cert}}$ ($k\ge2$) enters $\{V\le\ell\}$ within $\lceil V(\zeta_2)/\omega(\ell)\rceil$ updates and stays there.
\end{lem}
\begin{proof}
\emph{Decrease.} Since $\omega(0)=0$, a zero successor needs no argument. Otherwise put $\lambda=V(\zeta_{k+1})>0$ and $\chi=D(\lambda)^{-1}\zeta_{k+1}$. Then $S(\chi)=1$ and $\sigma_k\in\spow{\chi_1}{0}$. By \eqref{eq:normalized-recurrence} and \eqref{eq:weighted-homogeneity}, the scaled predecessor is
\begin{equation}
 \begin{aligned}
 \chi^-:={}&D(\lambda)^{-1}\zeta_k\\
 ={}&D(\lambda)^{-1}[D(\lambda)\chi-f(D(\lambda)\chi,\sigma_k,\delta_k)]\\
 ={}&\chi-\lambda^{-1}f(\chi,\sigma_k,\delta_k).
 \end{aligned}
 \label{eq:scaled-predecessor}
\end{equation}
Here $1/\lambda$ acts as a step size, unbounded at small levels: the predecessor is the successor moved against $f$, which points into $\mathcal K$, so convexity places it outside $\mathcal K$. Indeed, Lemma~\ref{cert:geometry} and \eqref{eq:certificate} give $S(\chi^-)\ge1-\lambda^{-1}\nabla S(\chi)^\top f(\chi,\sigma_k,\delta_k)\ge1+\epsilon/\lambda$, so $\chi^-\notin\mathcal K$ and $V(\chi^-)>1$ by \eqref{eq:sublevels}. It remains to turn this excess of $S$ into an excess of $V$. As $J$ is homogeneous of degree four along rays, $J(\chi')=S(\chi')J(\chi'/S(\chi')^{1/4})\le J_{\max}S(\chi')$ for $\chi'\ne0$, and the chain rule gives $\frac{\dd}{\dd\gamma}S(D(\gamma)^{-1}\zeta)=-J(D(\gamma)^{-1}\zeta)/\gamma$. Hence $G(\gamma)=S(D(\gamma)^{-1}\chi^-)$ satisfies $\gamma G'(\gamma)\ge-J_{\max}G(\gamma)$, so $\gamma^{J_{\max}}G(\gamma)$ is nondecreasing, and comparing $\gamma=1$ with $\gamma=V(\chi^-)$, where $G=1$, yields $V(\chi^-)^{J_{\max}}\ge S(\chi^-)$. With $V(\zeta_k)=\lambda V(\chi^-)$, $\ln(1+x)\ge x/(1+x)$ and $e^x\ge1+x$,
\begin{equation}
 \begin{aligned}
 V(\zeta_k)&\ge\lambda\exp\Bigl(\frac{\ln(1+\epsilon/\lambda)}{J_{\max}}\Bigr)\\
 &\ge\lambda\Bigl(1+\frac{\epsilon}{J_{\max}(\lambda+\epsilon)}\Bigr)=\lambda+\omega(\lambda).
 \end{aligned}
 \label{eq:implicit-decrease}
\end{equation}
For $\zeta_k=0$, \eqref{eq:uniform-decrement} forces $\zeta_{k+1}=0$.

\emph{Finite entry.} Since $\omega$ is increasing, every update whose successor lies outside $\{V\le\ell\}$ lowers $V$ by at least $\omega(\ell)>0$. If $n$ successors $\zeta_3,\dots,\zeta_{2+n}$ lay outside, then $\ell<V(\zeta_{2+n})\le V(\zeta_2)-n\,\omega(\ell)$, so $n<V(\zeta_2)/\omega(\ell)$; nonincrease prevents exit.
\end{proof}

\subsection{Local three-update capture}
\label{subsec:proof-capture}
As $\omega(\lambda)\le\lambda/J_{\max}$, \eqref{eq:uniform-decrement} alone allows $V$ to decay geometrically without reaching zero. Near the origin, however, every update takes the sliding branch of Proposition~\ref{prop:wellposed} and the recurrence becomes a closed-form map. For $0\le\Delta<k_3$, define
\begin{equation}
 \begin{gathered}
 L(\xi,y,r)=2|\xi|+|y|+|r|+k_2\sqrt{|\xi|},\\
 \mathcal C_\Delta=\{L\le k_3-\Delta\}.
 \end{gathered}
 \label{eq:local-capture-set}
\end{equation}
\begin{lem}[Three-update capture]
\label{lem:local-capture}
The set $\mathcal C_\Delta$ is an invariant neighborhood of zero for every sequence $|\delta_k|\le\Delta$. Its states reach zero in at most three updates.
\end{lem}
\begin{proof}
Continuity and $L(0)=0<k_3-\Delta$ give the neighborhood property. For $\Phi_1$ with inputs $(\xi,y,r+\delta)$, the selection variable is $s=(\xi+y+k_2\spow{\xi}{1/2}+r+\delta)/k_3$.
On $\mathcal C_\Delta$, its numerator has magnitude at most $L+\Delta\le k_3$. Proposition~\ref{prop:wellposed} therefore gives the sliding branch $\xi^+=0$, including $|s|=1$. The first component of \eqref{eq:normalized-recurrence} gives $y^+=-\xi$; the second then gives $r^+=-\xi-y-k_2\spow{\xi}{1/2}$. Thus the update is
\begin{equation}
 \mathcal T(\xi,y,r)=(0,-\xi,-\xi-y-k_2\spow{\xi}{1/2}).
 \label{eq:local-sliding-map}
\end{equation}
The third component of \eqref{eq:normalized-recurrence} selects $\sigma=(r+\delta-r^+)/k_3=s$, so the increment changes the sign selection, not the successor. Invariance follows from $L(\mathcal T(\xi,y,r))=|\xi|+|\xi+y+k_2\spow{\xi}{1/2}|\le2|\xi|+|y|+k_2\sqrt{|\xi|}\le L(\xi,y,r)$. Iterating gives
\begin{equation}
 \mathcal T^2(\xi,y,r)=(0,0,\xi),\qquad\mathcal T^3(\xi,y,r)=0.
 \label{eq:three-local-updates}\qedhere
\end{equation}
\end{proof}

\begin{thm}[Capture of the normalized recurrence]
\label{thm:capture-recurrence}
Let the gains be those of either design and $0\le\Delta\le\Delta_{\mathrm{cert}}$. For every $\zeta_2\in\R^3$ and every increment sequence satisfying $|\delta_k|\le\Delta$ for all $k\ge2$, the unique trajectory of \eqref{eq:normalized-recurrence} reaches zero at an index $k_{\mathrm{cap}}\le5+\lceil V(\zeta_2)/\omega(\ell)\rceil$ and remains there, with $V$, $\omega$ from Lemmas~\ref{lem:ilf} and~\ref{lem:decrease} and some $\ell>0$ depending only on the design and $\Delta$. The origin of \eqref{eq:normalized-recurrence} is Lyapunov stable uniformly over these sequences and the initial index, and therefore globally finite-step stable, that is, Lyapunov stable with finite-step convergence from every initial state.
\end{thm}
\begin{proof}[Proof of Theorems~\ref{thm:capture-recurrence} and~\ref{thm:capture}]
By Proposition~\ref{prop:certificate}, both designs satisfy the hypotheses of Lemmas~\ref{lem:ilf} and~\ref{lem:decrease}. Since $\Delta\le\Delta_{\mathrm{cert}}<k_3$, $\mathcal C_\Delta$ is a neighborhood of zero by Lemma~\ref{lem:local-capture}, and Lemma~\ref{lem:ilf} gives $\ell>0$, depending only on the design and $\Delta$, with $D(\ell)\mathcal K\subset\mathcal C_\Delta$. Lemma~\ref{lem:finite-entry} gives entry into it by the index $2+\lceil V(\zeta_2)/\omega(\ell)\rceil$, and at most three further updates reach zero by Lemma~\ref{lem:local-capture}, which gives the bound on $k_{\mathrm{cap}}$; origin invariance gives permanent capture.
Uniform Lyapunov stability follows from the continuity of $V$ at zero and its nonincrease along \eqref{eq:normalized-recurrence}, neither of which depends on the increment sequence or on the initial index (\appref{supp:capture}).
Section~\ref{subsec:transformed-recurrence} gives a finite $\zeta_2$ and increments obeying \eqref{eq:increment-bound}, so the recurrence result applies to the proposed controller, for $\Lambda\ne1$ after the scaling by $\Lambda$ described there. The normalization \eqref{eq:normalization} is invertible for fixed $T>0$, and zero normalized state is precisely \eqref{eq:transformed-capture}.
\end{proof}

\section{Third-order sampled-data accuracy}
\label{sec:accuracy}
This section proves Theorem~\ref{thm:accuracy}. Capture concerns the
controller states; it remains to bound their distance to the plant. As
the linear part of the loop under the input \eqref{eq:applied-input} is
nilpotent, this distance is an error that combines the
compensation errors of the last two sampling intervals only; after
capture, it is the plant state.

\subsection{Closed loop and error decomposition}
On each sampling interval, \eqref{eq:zoh-x1}--\eqref{eq:zoh-x2} remain valid if $u_k$
is replaced by $u_k+\phibar_{k-1}$ and $\phi$ by its deviation
$\phi-\phibar_{k-1}$ from the preceding average. Inserting
$u_k+\phibar_{k-1}$ from \eqref{eq:applied-input}
gives, for every $k\ge0$ and with $x_k=(x_{1,k},x_{2,k})^\top$,
\begin{gather}
 x_{k+1}=A_Tx_k+a_{k+1}\begin{pmatrix}1/2\\1/T\end{pmatrix}+g_k,
 \label{eq:closed-loop}\ifextended\else\displaybreak[0]\fi\\
 A_T=\begin{pmatrix}1/2&T/4\\-1/T&-1/2\end{pmatrix},\notag\\
 g_k=\int_0^T\begin{pmatrix}T-\tau\\1\end{pmatrix}
     \bigl[\phi(t_k+\tau)-\phibar_{k-1}\bigr]\dd\tau.\notag
\end{gather}
The matrix $A_T$ is nilpotent, $A_T^2=0$, as for the two-step deadbeat
law of Section~\ref{subsec:deadbeat}; the proper implicit
super-twisting controller likewise becomes deadbeat near its
origin~\citep[Remark~3.3, Theorem~3.4]{SeeberAndritsch2025}. The term
$g_k$ is the effect of the compensation error over one sampling interval.

Without this error, the plant would follow the controller states, with
$a_k$ in the role of $c_k$ and $b_k$ in that of $x_{2,k}$. Indeed,
$A_T(a_k-Tb_k/2,\,b_k)^\top=(a_k/2,\,-a_k/T)^\top$, so by
\eqref{eq:retained-difference} the pair $(a_k-Tb_k/2,\,b_k)$
satisfies \eqref{eq:closed-loop} with $g_k=0$, and by \eqref{eq:startup}
and \eqref{eq:measured-combination} it equals $x(0)$ at $k=0$. The
difference $e_k$ between $x_k$ and this pair therefore satisfies
$e_{k+1}=A_Te_k+g_k$ and $e_0=0$, so $e_1=g_0$ and, as $A_T^2=0$,
\begin{equation}
 \begin{gathered}
 x_{1,k}=a_k-\frac T2b_k+e_{1,k},\qquad x_{2,k}=b_k+e_{2,k},\\
 e_k=g_{k-1}+A_Tg_{k-2}\quad(k\ge2).
 \end{gathered}
 \label{eq:physical-identities}
\end{equation}
Capture gives $a_k=b_k=0$ and hence $x_k=e_k$ for $k\ge k_{\mathrm{cap}}$.
At $k=2$, however,
$e_2=g_1+A_Tg_0$ involves the initial value $\phibar_{-1}=0$ in place of
an average: for constant $\phi=d$, where $\Delta=0$,
$e_2=A_Tg_0=(T^2d/2,\,-Td)^\top$. Bounds proportional to $\Delta$ thus
start at $k=3$, hence $k_{\mathrm{acc}}=\max\{k_{\mathrm{cap}},3\}$.

\subsection{Proof of the accuracy bounds}
\label{subsec:accuracy-proof}
\begin{proof}[Proof of Theorem~\ref{thm:accuracy}]
\emph{Sampled bounds.} For $j\ge1$ and $\tau,s\in[0,T]$,
\eqref{eq:disturbance-class} gives
$|\phi(t_j+\tau)-\phi(t_{j-1}+s)|\le\Delta(T+\tau-s)$, and averaging over
$s$ gives $|\phi(t_j+\tau)-\phibar_{j-1}|\le\Delta(\tau+T/2)$. With
$\int_0^T(T-\tau)(\tau+T/2)\dd\tau=5T^3/12$ and
$\int_0^T(\tau+T/2)\dd\tau=T^2$, the components of
$g_j=(g_{1,j},g_{2,j})^\top$ in \eqref{eq:closed-loop} satisfy
$|g_{1,j}|\le5\Delta T^3/12$ and $|g_{2,j}|\le\Delta T^2$. For $k\ge3$,
these bounds hold for $g_{k-1}$ and $g_{k-2}$, and
$e_k=g_{k-1}+A_Tg_{k-2}$ gives
\begin{align}
 |e_{1,k}|&\le|g_{1,k-1}|+\frac{|g_{1,k-2}|}{2}+\frac{T|g_{2,k-2}|}{4}\notag\\
  &\le\Bigl(\frac5{12}+\frac5{24}+\frac14\Bigr)\Delta T^3=C_1\Delta T^3,\notag\displaybreak[0]\\
 |e_{2,k}|&\le|g_{2,k-1}|+\frac{|g_{1,k-2}|}{T}+\frac{|g_{2,k-2}|}{2}\notag\\
  &\le\Bigl(1+\frac5{12}+\frac12\Bigr)\Delta T^2=C_2\Delta T^2,
 \label{eq:error-bound}
\end{align}
and $x_k=e_k$ for $k\ge k_{\mathrm{acc}}$.

\emph{Attained bound.} The ramp satisfies \eqref{eq:disturbance-class}, so
Theorem~\ref{thm:capture} applies, and
$\phi(t_j+\tau)-\phibar_{j-1}=\Delta(\tau+T/2)$ for $j\ge1$. Hence
$g_j=(5\Delta T^3/12,\,\Delta T^2)^\top$, the three terms of $e_{1,k}$ are
positive, and $x_{1,k}=e_{1,k}=C_1\Delta T^3$ for $k\ge k_{\mathrm{acc}}$.
The same values give $x_{2,k}=(1-5/12-1/2)\Delta T^2=\Delta T^2/12$.

\emph{Between samples.} For $k\ge k_{\mathrm{acc}}$, the samples at both
ends of $[t_k,t_{k+1}]$ obey \eqref{eq:sampled-bounds}. On this interval,
$\dot x_2=u_k+\phi$ has the average $(x_{2,k+1}-x_{2,k})/T$, at most
$2C_2\Delta T$ in magnitude, from which it differs by at most
$\Delta T/2$; hence $|u(t)+\phi(t)|\le C_3\Delta T$. A function whose
second derivative is bounded by $M$ deviates from its chord over the
interval by at most $MT^2/8$; with $M=\Delta$ for $x_2$ and
$M=C_3\Delta T$ for $x_1$, this gives \eqref{eq:intersample-bounds}
(\appref{supp:accuracy}).

\emph{Rejection.} For $k\ge k_*$, $[t_{k-3},t_k]\subset[t_*,\infty)$, so
the deviations $\phi-\phibar$ in $g_{k-1}$ and $g_{k-2}$ vanish and
$x_k=e_k=0$ by \eqref{eq:physical-identities}.
As $x_k=0$ and $a_{k+1}=0$, \eqref{eq:applied-input} gives
$u_k=-\phibar_{k-1}=-d$, so $\dot x_2=0$ on $[t_k,t_{k+1}]$ and the plant
stays at zero. Finally, $q_k=0$ gives $v_k=-\phibar_{k-2}=-d$.
\end{proof}
The constants $C_2$ and $C_3$ and the intersample constants are not
claimed to be sharp.

\begin{figure*}[t]
\centering
\includegraphics[width=0.80\textwidth]{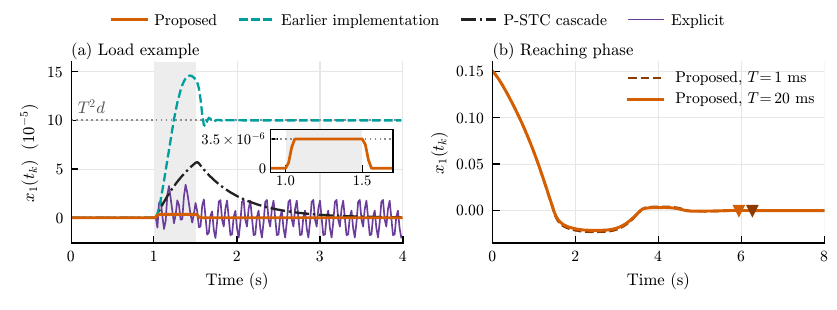}
\caption{Sampled output. (a) Load example, $T=20$~ms: the load rises during the shaded interval and then stays constant; dotted: offset $T^2d$ \eqref{eq:cdc-offset}; inset: proposed controller with its bound $C_1\Delta T^3$ (dotted). (b) Reaching phase of the proposed controller for $T=1$~ms and $T=20$~ms; triangles: capture.}
\label{fig:sim-load}
\end{figure*}

\begin{figure*}[t]
\centering
\includegraphics[width=0.80\textwidth]{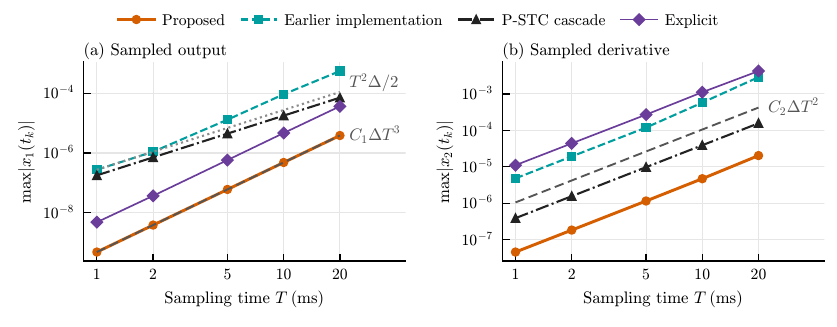}
\caption{Largest sampled errors over $12\le t_k<16$~s for the disturbance and the initial state of the reaching phase (design A). Gray dashed: bounds $C_1\Delta T^3$ and $C_2\Delta T^2$ of Theorem~\ref{thm:accuracy}; gray dotted: $T^2\Delta/2$, the offset \eqref{eq:cdc-offset} at the peak of the disturbance.}
\label{fig:sim-orders}
\end{figure*}

\subsection{Role of the implicit step and small sampling times}
\label{subsec:small-period}
The derivation of \eqref{eq:physical-identities} uses only the input
\eqref{eq:applied-input}, the initialization \eqref{eq:startup} and
$b_{k+1}=(a_{k+1}-a_k)/T$. The bounds of Theorem~\ref{thm:accuracy}
therefore hold for every sequence $a_k$ from the index $k\ge3$ from
which $a_k=b_k=0$: the orders $(3,2)$ are those of the deadbeat law
\eqref{eq:applied-input} with its one-step-delayed compensation on the
exact sampled model and are not inherited from homogeneity. The implicit
PID-like step decides how this index is reached. Feeding the average
into the step \eqref{eq:proposed-step} lets the integral state cancel
the disturbance, as in the continuous controller, so that the input
keeps the PID-like form \eqref{eq:pid-corrections} without a separate
feedforward term; the price is the condition on $\Delta$.

Deadbeat control from $k=0$,
$a_{k+1}\equiv0$, has the same accuracy but the first input
$u_0=-x_{1,0}/T^2-3x_{2,0}/(2T)$, in general of order $T^{-2}$\ifextended\ (\appref{supp:remarks})\fi.
For a fixed initial condition and disturbance,
the capture time and the compensated input $u_k+\phibar_{k-1}$ of the
proposed controller instead stay bounded as $T\downarrow0$. For
$\Lambda=1$, $\zeta_2=D(T)^{-1}(a_2,b_2,q_2)^\top$ by
\eqref{eq:normalization}, so $V(\zeta_2)=V(a_2,b_2,q_2)/T$ by
Lemma~\ref{lem:ilf}, and Theorem~\ref{thm:capture-recurrence} gives
$t_{k_{\mathrm{cap}}}<6T+V(a_2,b_2,q_2)/\omega(\ell)$.
\begin{cor}[Small sampling times]
\label{cor:small-period}
Under the hypotheses of Theorem~\ref{thm:capture} with $\Lambda=1$, fix
$x(0)$, $v_0$ and
one disturbance satisfying \eqref{eq:disturbance-class}, and let
$z_0=(x_{1,0},x_{2,0},v_0+\phi(0))$. As $T\downarrow0$,
$(a_2,b_2,q_2)\to z_0$ and
$\limsup_{T\downarrow0}t_{k_{\mathrm{cap}}}\le V(z_0)/\omega(\ell)$. For
every $T>0$ and $k\ge3$,
$|u_k+\phibar_{k-1}|\le c_{\mathcal K}V(a_2,b_2,q_2)+(C_1+3C_2/2)\Delta T$,
where
\begin{equation*}
c_{\mathcal K}=\max_{\chi\in\mathcal K}(k_1|\chi_1|^{1/3}+k_2|\chi_2|^{1/2}+|\chi_3|)
\end{equation*}
depends only on the design; this bound tends to $c_{\mathcal K}V(z_0)$,
and $u_0,u_1,u_2$ stay bounded as $T\downarrow0$.
\end{cor}
For the proof, \eqref{eq:physical-identities} turns
\eqref{eq:applied-input} into the law \eqref{eq:continuous-pid} at
$(a_{k+1},b_{k+1},q_{k+1})$ plus a correction that
\eqref{eq:error-bound} bounds, and $V(a_k,b_k,q_k)=TV(\zeta_k)$ does not
increase (\appref{supp:accuracy}).

\section{Simulation study}
\label{sec:simulations}
Three experiments illustrate the rejection of a constant load, the
reaching phase and the accuracy orders. The proposed controller is
compared with the
earlier implementation \eqref{eq:cdc-implementation}, with the explicit
(forward-Euler) discretization of \eqref{eq:continuous-pid} and with
the proportional--super-twisting (P-STC) cascade of \citet{Zapf2025},
in which the modified implicit super-twisting controller of
\citet{Andritsch2024} with the gains $k_2$ and $k_3$ acts on
$x_{2,k}+2x_{1,k}$. Both designs have $\Lambda=1$; the runs use design A
unless stated otherwise, and every integral state starts at zero. Every
run integrates its plant exactly between the samples; off the sliding
branch, Brent's method~\citep{Brent1973} solves
\eqref{eq:scalar-residual} to a residual of at most
$3.6\times10^{-6}\,T^2$.
\appref{supp:simulations} gives the comparators' laws and the tested
values.

\emph{Load disturbance.}\label{subsec:sim-load} The proposed controller
rejects a constant load exactly after three samples, where the
comparators show an offset, a slow decay or chattering
(Figure~\ref{fig:sim-load}(a)). The plant starts at rest, $T=20$~ms, and
from $t=1$~s the load rises with the slope
$\Delta=0.5\le\Delta_{\rm cert}$ to
$d=0.25$ and then stays constant. The output of the proposed controller
reaches the bound $C_1\Delta T^3=3.5\times10^{-6}$ while the load rises
and is zero from the third sample after the load has become constant,
as Theorem~\ref{thm:accuracy}(iii) states; its input stays within
$\Delta T=0.01$ of the equivalent control $-\phi$. The earlier
implementation settles, without chattering, at its offset
$T^2d=10^{-4}$, and the error of the cascade decays slowly. The explicit
discretization shows discretization chattering: while the load is
constant, its input alternates about $-d$ at every sample with
amplitudes of $0.19$ to $0.37$, and its output error reaches \mbox{$2.1\times10^{-5}$}.

\emph{Reaching phase.} The transient from a nonzero initial state
changes little with the sampling time (Figure~\ref{fig:sim-load}(b));
the load example never leaves the sliding branch, where the gains play
no role. For $x(0)=(0.15,-0.04)$
and $\phi=(\Delta/2)\sin(2t)$ with $\Delta=\Delta_{\rm cert}$, the
sampled outputs for $T=20$~ms and $T=1$~ms nearly coincide. Capture,
taken as the first sample from which $|a_k|,|b_k|,|q_k|\le10^{-12}$,
occurs at $5.94$~s and $6.27$~s, respectively, and the largest inputs are $0.62$ and
$0.67$, in line with Corollary~\ref{cor:small-period}, whose bound on
the capture time, $1.25\times10^4$~s, is conservative
(\appref{supp:capture}). Deadbeat control from $k=0$ has
the same accuracy (Section~\ref{subsec:small-period}) but needs inputs
up to $374$ and \mbox{$1.5\times10^5$}.

\emph{Accuracy order.}\label{subsec:sim-orders} After capture, the errors
scale with the sampling time as Theorem~\ref{thm:accuracy} states
(Figure~\ref{fig:sim-orders}). For the disturbance and the initial state
of the reaching phase and $T\in\{1,2,5,10,20\}$~ms, the largest sampled
errors of the proposed controller have the least-squares slopes $3.00$
for $x_1$ and $2.03$ for $x_2$ in the log-log plot, as they have for
design B with $\Delta=1$. Those of $x_1$ equal the
bound $C_1\Delta T^3$ within $0.03\%$; those of $x_2$ stay below $5\%$
of $C_2\Delta T^2$. The explicit discretization has the same orders in
$T$, with output errors about ten times larger. The errors of the
cascade are of second order, and the output errors of the earlier
implementation approach its offset $T^2\Delta/2$ as $T$ decreases.

\section{Model errors and measurement noise}
\label{subsec:sim-limits}
On the sliding branch, \eqref{eq:applied-input} is the deadbeat law of
Section~\ref{subsec:deadbeat}, and for $\phi=0$ the stability of the
linear loop depends on the input-gain ratio, lag or delay relative
to $T$, not on the gains; only this branch is analyzed.

\emph{Input-gain error, lag and delay.} If the plant receives $\beta u_k$,
\eqref{eq:reconstruct-average} returns $\phibar_{k-1}+(\beta-1)u_{k-1}$,
and the characteristic polynomial $\beta z^3+(1-\beta)(z-1)^3$ of the
loop is Schur stable exactly for $1/2<\beta<8/7$.
For a first-order actuator lag
$\tau_{\rm a}\dot u_{\rm a}=u-u_{\rm a}$ or an input delay
$\tau_{\rm d}\le T$, during which the previous input is still applied,
\eqref{eq:reconstruct-average} uses the commanded input. Computed
spectral radii then give stability for $\tau_{\rm a}<0.1538\,T$ and
$\tau_{\rm d}<0.150\,T$, instability at all larger tested values.

\emph{Simulations.} Simulations of the load example at six sampling
times from $1$ to $50$~ms agree with these limits
(\appref{supp:simulations}). For every $T$, the proposed controller
stays on the sliding branch for every tested $\beta$ in $(1/2,8/7)$
and for none outside $[1/2,8/7]$, and for lags and delays up to $0.14\,T$;
for larger ones it leaves the branch or diverges. With a lag or delay
of $0.1\,T$ at $20$~ms, its output error stays below
$4.5\times10^{-6}$, against $3.5\times10^{-6}$ without one. Under the
common measurement-to-input delay of one sampling time, however, the
proposed controller diverges at every tested $T$, whereas the earlier
implementation and the cascade stay bounded, as they do for lags up to
$2\,T$ and for $\beta\in\{0.3,0.5,2,3\}$.

\emph{Measurement noise.}
Noise $n_{1,k}$ on $x_1$ adds $-n_{1,k}/T^2$ to the input
\eqref{eq:applied-input}: at $T=1$~ms, noise of $10^{-6}$ changes the
input by~$1$, four times the load of Figure~\ref{fig:sim-load}(a).
Noise $|n_{2,k}|\le\nu_2$ on $x_2$ adds at most $4\nu_2/T^2$ to
$\delta_k$ through \eqref{eq:reconstruct-average}; if
$\Delta+4\nu_2/T^2\le\Delta_{\mathrm{cert}}$, the controller states are
still captured in the sense of Theorem~\ref{thm:capture}, with the
reconstructed averages in place of $\phibar$ in $q_k$.
Theorem~\ref{thm:accuracy} does not carry over to noisy measurements.

\section{Conclusions}
\label{sec:conclusions}
A digital PID-like sliding-mode controller has been developed for a
double integrator with a
Lipschitz matched disturbance. Its proper implicit discretization is
matched to the exact sampled dynamics of the plant: the implicit step
acts on the controller's own states, and the output extrapolated by half
a sampling time and a causal reconstruction of the disturbance average
make the plant follow them. This removes the output offset of the
earlier implicit implementation.

For two certified gain designs and every sampling time, the
controller states reach a discrete-time sliding mode after a finite
number of steps; the proof uses an implicit Lyapunov function and a
computer-assisted certificate of its decrease.
In this sliding mode, each sample of the plant state depends only on the
disturbance over the last three sampling intervals.
The closed loop then has third-order output accuracy and
second-order accuracy for the derivative, rejects constant disturbances
exactly after a finite transient, and is free of discretization
chattering.
The digital controller thus attains on the sampled-data plant the
accuracy orders of the continuous-time PID-like controller, and no
deterministic causal controller using the same sampled information
improves them.

\section*{\raggedright Declaration of generative AI and AI-assisted technologies in the manuscript preparation process}
During the preparation of this work, the authors used Claude Opus~5.5, Claude Opus~5 and Claude Fable~5
(Anthropic) to assist with software development, with checking derivations
and with drafting and refining the manuscript. After using these tools,
the authors reviewed and edited the content and take full responsibility
for the content of the published article.

\ifextended
\appendix
\input{appendix.tex}
\fi

\bibliographystyle{elsarticle-harv}
\bibliography{refs}

\end{document}

%% file: appendix.tex
\section{Proof of the certificate by exact interval arithmetic}
\label{app:certificate}

Proposition~\ref{prop:certificate} asserts two inequalities at every point of the unit level set $\{S=1\}$, for every admissible sign selection and increment; Lemma~\ref{lem:ilf} needs the second to define $V$, and Lemma~\ref{lem:decrease} the first for its decrease. Checking finitely many sample points does not establish the inequalities between them. Interval arithmetic instead computes, for a whole rectangle of parameters, an interval that contains every value of the quantity to be bounded; if this interval meets the required margin, the inequality holds on the entire rectangle. Finitely many such rectangles covering a parameterization of $\{S=1\}$ prove the inequalities everywhere. All arithmetic is exact and rational; only roots are approximated, with outward rounding.

The proof uses only Lemmas~\ref{cert:geometry} and~\ref{supp:scaling-identities}, the definitions of $f$ in \eqref{eq:vector-field} and of $S$ and $J$ in \eqref{eq:quartic}, the gains and $\Delta_{\mathrm{cert}}$ of Table~\ref{tab:certified-gains} for $\Lambda=1$, and the matrices of Table~\ref{tab:certificate-matrices}; no Lyapunov or capture conclusion is used. The matrices were found by a floating-point search that imposed both inequalities at sampled boundary points and maximized first the slope bound and then, at $\Delta_{\mathrm{cert}}=k_3/2$, the inward margin; they were then rounded to rationals, and only the rounded matrices enter the proof. The program that performs the computations, called the verifier below, first checks that each matrix is symmetric and has positive leading principal minors, which is equivalent to symmetric positive definiteness (SPD), the hypothesis of Lemma~\ref{cert:geometry}.

The proof has five steps. (i)~Radial scaling gives the point of $\{S=1\}$ on each ray from the origin, and both quantities of \eqref{eq:certificate} at this point, in closed form (Section~\ref{cert:normalization-section}). (ii)~Each ray with $\chi_1\ge0$ meets the surface of a box around the origin once. The half of the surface met by these rays is cut into twelve rectangular patches, each parameterized by a unit square in coordinates that turn the cube and square roots in $f$ into polynomials (Section~\ref{cert:coverage-section}). (iii)~Both quantities are affine in $(\sigma,\delta)$ and unchanged when the signs of $(\chi,\sigma,\delta)$ are reversed, so $\sigma=1$ and $\delta=\pm\Delta_{\mathrm{cert}}$ suffice (Proposition~\ref{cert:reduction}). (iv)~Interval arithmetic bounds the resulting functions on subrectangles of these squares (Section~\ref{interval:evaluation}). (v)~Bisection refines the subrectangles until each meets the margins, which proves Proposition~\ref{prop:certificate} (Section~\ref{interval:cover}).

\begin{figure}[t]
\centering
\begin{tikzpicture}[scale=1.05,>=stealth,font=\small]
 \draw[->,gray] (-0.25,0) -- (3.9,0) node[below] {$\eta_1$};
 \draw[->,gray] (0,-2.35) -- (0,2.45) node[left] {$\eta_2$};
 \node[anchor=north east,inner sep=2pt] at (0,0) {$0$};
 \fill[gray!25] (1.7,0.45) -- (3.024,0.8005)
   arc[start angle=23.59,end angle=42.68,x radius=3.3,y radius=2] -- (1.7,0.95) -- cycle;
 \draw[thick] (0,-2) arc[start angle=-90,end angle=90,x radius=3.3,y radius=2];
 \node[anchor=west] at (2.7,-1.5) {$\{S=1\}$};
 \draw[very thick] (0,1.25) -- (1.7,1.25);
 \draw[very thick,densely dashed] (1.7,1.25) -- (1.7,0);
 \draw[very thick] (1.7,0) -- (1.7,-1.25);
 \draw[very thick,densely dashed] (1.7,-1.25) -- (0,-1.25);
 \foreach \y in {1.25,0,-1.25} \fill (1.7,\y) circle (1.1pt);
 \node[anchor=south] at (0.85,1.25) {$h_2=1$};
 \node[anchor=north] at (0.85,-1.25) {$h_2=1$};
 \node[anchor=east,inner sep=3pt] at (1.7,-0.62) {$h_1=1$};
 \node[anchor=east] at (-0.1,0.62) {$s_2=1$};
 \node[anchor=east] at (-0.1,-0.62) {$s_2=-1$};
 \draw[line width=2.2pt] (1.7,0.45) -- (1.7,0.95);
 \draw[line width=2.2pt] (3.024,0.8005)
   arc[start angle=23.59,end angle=42.68,x radius=3.3,y radius=2];
 \node[anchor=west,inner sep=3pt] at (1.7,0.52) {$\mathbf B$};
 \draw[->] (0,0) -- (1.7,0.7);
 \draw[densely dotted] (1.7,0.7) -- (2.7296,1.124);
 \fill (1.7,0.7) circle (1.3pt);
 \node[anchor=south east,inner sep=2pt] at (1.66,0.74) {$\eta(h)$};
 \fill (2.7296,1.124) circle (1.3pt);
 \node[anchor=south west,inner sep=2pt] at (2.76,1.15) {$\chi=\eta/S_\eta^{1/4}$};
\end{tikzpicture}
\caption{The construction of Appendix~\ref{app:certificate}, drawn schematically in the plane $\eta_3=0$. Each ray from the origin with $\eta_1\ge0$ meets the surface of a box at one point $\eta(h)$ and the level set $\{S=1\}$ at $\chi=\eta/S_\eta^{1/4}$ (Lemma~\ref{cert:normalization}). The planes $\eta_2=0$ and $\eta_3=0$ cut the half $\eta_1\ge0$ of the box surface into twelve rectangular patches, eight of which meet this plane, in the four edges drawn alternately solid and dashed; each is the image of a unit square of parameters $h$ (Lemma~\ref{cert:coverage}). The verifier bounds the two quantities of \eqref{eq:certificate} on the piece of $\{S=1\}$ (thick arc) that belongs to a rectangle $\mathbf B$ of parameters.}
\label{fig:patches}
\end{figure}
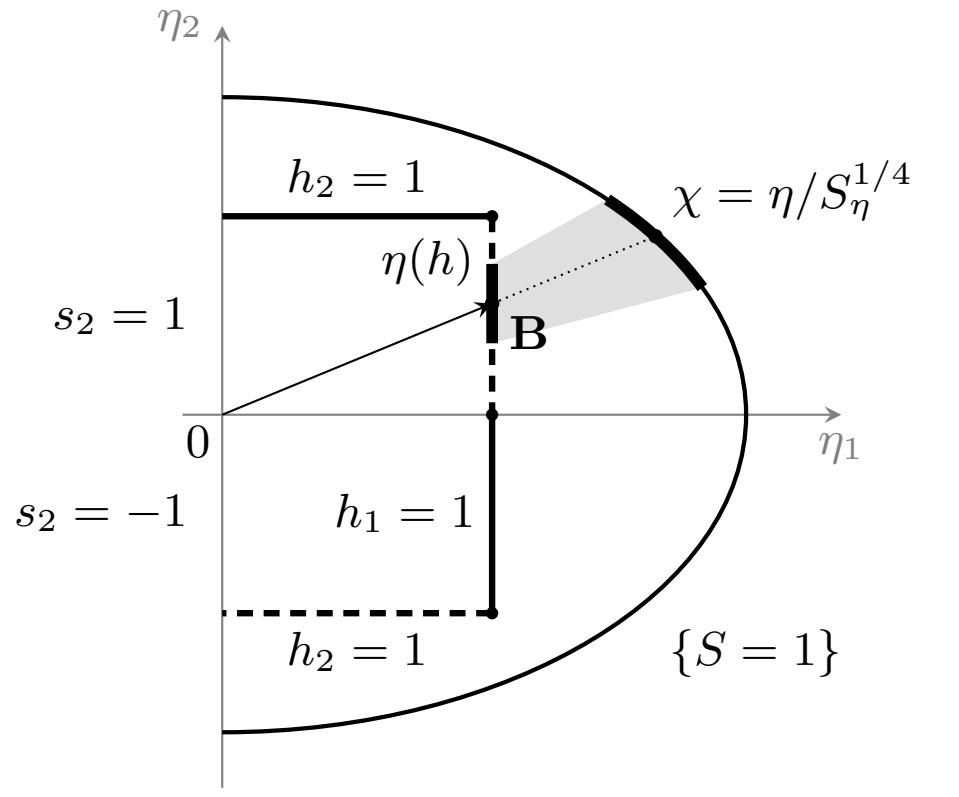

A reader who wants to know what is proved, and not how the program works, needs only three statements. Lemma~\ref{cert:normalization} writes both quantities of \eqref{eq:certificate} as explicit functions of a ray direction. Proposition~\ref{cert:reduction} states the finitely many families of inequalities that suffice. Proposition~\ref{interval:uniform} states that a completed run of the verifier establishes them. Sections~\ref{interval:evaluation} and~\ref{interval:cover} specify the computation in enough detail to reimplement it. Figure~\ref{fig:patches} shows the construction in a plane. The symbols $q_i$, $g$, $H$, $h$, $s_2$, $s_3$ and $\rho_j$ are local to this section and unrelated to $q_k$, $g_k$ and $s$ of the main text, to $h$ of Appendix~\ref{supp:accuracy} and to $\rho$ of Appendix~\ref{supp:simulations}.

The identities used below are collected first; \eqref{cert:radial} is needed only in Appendix~\ref{supp:proofs}.
\begin{lem}[Scaling identities]
\label{supp:scaling-identities}
For every $\chi\in\R^3$ and real $t$,
\begin{equation}
 \begin{gathered}
 \nabla S(\chi)=4\sum_i(\chi^\top P_i\chi)P_i\chi,\quad
 S(t\chi)=t^4S(\chi),\\
 \nabla S(t\chi)=t^3\nabla S(\chi),\quad J(t\chi)=t^4J(\chi).
 \end{gathered}
 \label{cert:gradient-general}
\end{equation}
For $\chi\ne0$, the radial normalization $\widehat\chi=\chi/S(\chi)^{1/4}$ satisfies $S(\widehat\chi)=1$ and $J(\chi)=S(\chi)J(\widehat\chi)$. For every $\chi,\zeta\in\R^3$ and $\gamma>0$,
\begin{equation}
 \begin{gathered}
 \frac{\dd}{\dd\gamma}S(D(\gamma)\chi)=\frac{J(D(\gamma)\chi)}\gamma,\\
 \frac{\dd}{\dd\gamma}S(D(\gamma)^{-1}\zeta)=-\frac{J(D(\gamma)^{-1}\zeta)}\gamma.
 \end{gathered}
 \label{cert:radial}
\end{equation}
\end{lem}
\begin{proof}
The chain rule gives the gradient, and homogeneity of each quadratic gives the scaling identities in \eqref{cert:gradient-general}. For $\chi\ne0$, positive definiteness gives $S(\chi)>0$, and $t=S(\chi)^{-1/4}$ in \eqref{cert:gradient-general} gives $S(\widehat\chi)=1$ and $J(\widehat\chi)=J(\chi)/S(\chi)$. Finally, $\frac{\dd}{\dd\gamma}D(\gamma)^{\pm1}=\pm WD(\gamma)^{\pm1}/\gamma$ with the chain rule gives \eqref{cert:radial}.
\end{proof}

\begin{table*}[t]
\centering
\caption{Certificate matrices of both designs, exact: the six independent entries of each symmetric matrix, $(1,1)$, $(1,2)$, $(1,3)$ in its first line and $(2,2)$, $(2,3)$, $(3,3)$ in its second.}
\label{tab:certificate-matrices}
\begin{tabular}{llrrr}
\toprule
A & $P_1$ & $143186595/80950024$ & $79435511/99171693$ & $137565/1189136$\\
 & & $10570314/29058569$ & $3216769/61641123$ & $107921/13976665$\\
 & $P_2$ & $279014956/91746283$ & $230708769/84701564$ & $-64055437/88563575$\\
 & & $246777171/77661710$ & $-29064249/45612334$ & $14972877/86909005$\\
 & $P_3$ & $41296326/45032401$ & $-57988620/38331287$ & $4758066/88972069$\\
 & & $30755446/12318233$ & $-5807976/65745899$ & $208607/64215006$\\
\midrule
B & $P_1$ & $2293365850/45379223$ & $720846837/51534116$ & $9228894/59214061$\\
 & & $461299807/96148754$ & $-11479780/92885043$ & $953311/31136877$\\
 & $P_2$ & $816546599/39283333$ & $-573571124/31448555$ & $19200733/90922961$\\
 & & $798757558/49868391$ & $-15374412/82686757$ & $82749/35994148$\\
 & $P_3$ & $3813819041/54174173$ & $1714218229/72347612$ & $-200744670/73748039$\\
 & & $1483853309/74244811$ & $-32691703/31091649$ & $5234533/48967396$\\
\bottomrule
\end{tabular}
\end{table*}

\subsection{Radial normalization}
\label{cert:normalization-section}

Proposition~\ref{prop:certificate} concerns individual points of $\{S=1\}$, so any parameterization of this surface will do. Radial scaling along straight rays, as in the proof of Lemma~\ref{lem:ilf}, finds the point of $\{S=1\}$ on each ray in closed form, whereas the weighted dilation $D(\gamma)$ would require solving \eqref{eq:ilf} for every point. The functions $g$ and $H$ defined next are, up to the factor $4/S_\eta$, the two quantities of \eqref{eq:certificate} at the point of $\{S=1\}$ on the ray of $\eta$. For $\eta\in\R^3$ with $\eta\ne0$ and $\eta_1\ge0$, define
\begin{equation}
 \begin{gathered}
 q_i(\eta)=\eta^\top P_i\eta,\qquad
 N(\eta)=\sum_{i=1}^3 q_i(\eta)\,P_i\eta,\\
 S_\eta=S(\eta)=\sum_{i=1}^3 q_i(\eta)^2,
 \end{gathered}
 \label{cert:raydata}
\end{equation}
and, for arbitrary real $\sigma,\delta$,
\begin{equation}
 \begin{aligned}
 g(\eta,\sigma,\delta)={}&N_1\eta_2
 +N_2\Big[-k_1\eta_1^{1/3}S_\eta^{1/6}\\
 &-k_2\spow{\eta_2}{1/2}S_\eta^{1/8}+\eta_3\Big]\\
 &+N_3(-k_3\sigma+\delta)\,S_\eta^{1/4},\\
 H(\eta)={}&N(\eta)^\top W\eta,
 \end{aligned}
 \label{cert:gH}
\end{equation}
where $N_j$ is the $j$-th coordinate of $N(\eta)$ and $\eta_1^{1/3}$ is the ordinary cube root of $\eta_1\ge0$. The next lemma is step~(i).
\begin{lem}[Radial normalization]
\label{cert:normalization}
Let $\eta$ be as above. Then $S_\eta>0$, the unique point of the ray of $\eta$ on $\{S=1\}$ is $\chi=\eta/S_\eta^{1/4}$, $\spow{\chi_1}{0}=\spow{\eta_1}{0}$, and
\begin{equation}
 \nabla S(\chi)^\top f(\chi,\sigma,\delta)=\frac{4\,g(\eta,\sigma,\delta)}{S_\eta},
 \quad
 J(\chi)=\frac{4\,H(\eta)}{S_\eta}.
 \label{cert:identities}
\end{equation}
\end{lem}

\begin{proof}
Positive definiteness gives $S_\eta>0$. By \eqref{cert:gradient-general}, $S(t\eta)=t^4S_\eta$, so $t=S_\eta^{-1/4}$ is the only $t>0$ with $S(t\eta)=1$, and scaling by it preserves coordinate signs. At $\chi$, the gradient and the state terms of $f$ become
\begin{equation}
 \begin{aligned}
 \nabla S(\chi)&=4\sum_i\frac{q_i(\eta)P_i\eta}{S_\eta^{1/2+1/4}}
 =\frac{4N(\eta)}{S_\eta^{3/4}},\\
 \spow{\chi_1}{1/3}&=\eta_1^{1/3}S_\eta^{-1/12},\qquad
 \chi_j=\eta_jS_\eta^{-1/4},\\
 \spow{\chi_2}{1/2}&=\spow{\eta_2}{1/2}S_\eta^{-1/8}.
 \end{aligned}
 \label{cert:gradient}
\end{equation}
Multiplying the scalar product by $S_\eta/4$ leaves the five terms
\begin{equation}
 \begin{aligned}
 &\frac{S_\eta}{4}\nabla S(\chi)^\top f(\chi,\sigma,\delta)\\
 &\quad=N_1\eta_2-k_1N_2\eta_1^{1/3}S_\eta^{1-3/4-1/12}\\
 &\qquad-k_2N_2\spow{\eta_2}{1/2}S_\eta^{1-3/4-1/8}+N_2\eta_3\\
 &\qquad+N_3(-k_3\sigma+\delta)S_\eta^{1-3/4}=g(\eta,\sigma,\delta).
 \end{aligned}
 \label{cert:expanded}
\end{equation}
The linear state terms have exponent $1-3/4-1/4=0$. Similarly,
$J(\chi)=4N(\eta)^\top W\eta/S_\eta^{3/4+1/4}=4H(\eta)/S_\eta$.
\end{proof}

\subsection{Closed patches and all sign selections}
\label{cert:coverage-section}

The patches below are step~(ii). Fix positive $\rho_1,\rho_2,\rho_3$ and consider the box $|\eta_1|\le\rho_1^3$, $|\eta_2|\le\rho_2^2$, $|\eta_3|\le\rho_3$. The planes $\eta_2=0$ and $\eta_3=0$ cut the half $\eta_1\ge0$ of its surface into twelve rectangles, the patches (Figure~\ref{fig:patches}). They are parameterized so that the cube and square roots in $f$ become polynomials: for parameters $h=(h_1,h_2,h_3)\in[0,1]^3$ and patch signs $s_2,s_3\in\{-1,1\}$ put
\begin{equation}
 \begin{gathered}
 \eta(h;s_2,s_3)=
 \big(\rho_1^3h_1^3,\ \ s_2\rho_2^2h_2^2,\ \ s_3\rho_3h_3\big)^\top,\\
 \max_jh_j=1,
 \end{gathered}
 \label{cert:patches}
\end{equation}
so that, also for vanishing parameters,
\begin{equation}
 \eta_1^{1/3}=\rho_1h_1,\qquad \spow{\eta_2}{1/2}=s_2\rho_2h_2 .
 \label{cert:root-removal}
\end{equation}
Fixing one $h_j=1$ leaves two parameters in $[0,1]^2$. The three choices of $j$ and four sign pairs give twelve closed patches, all with $\eta\ne0$ and $\eta_1\ge0$. The verifier uses $\rho=(1/2,1/2,2)$ for design A and $\rho=(1/3,1/2,3)$ for design B; Lemma~\ref{cert:coverage} holds for all positive $\rho_j$, so these values affect only the computational effort.
On a patch, \eqref{cert:root-removal} replaces $\eta_1^{1/3}$ by $\rho_1h_1$ and $\spow{\eta_2}{1/2}$ by $s_2\rho_2h_2$ in \eqref{cert:gH}, so that $q_i(\eta)$, $N(\eta)$, $S_\eta$ and $H(\eta)$ are polynomials in the two free parameters and the only non-polynomial ingredients of $g$ are the three roots of the positive polynomial $S_\eta$.

\begin{lem}[Coverage of the half-boundary]
\label{cert:coverage}
The twelve normalized patch images cover the half-boundary exactly: their union is \mbox{$\{\chi:S(\chi)=1,\ \chi_1\ge0\}$}.
\end{lem}
\begin{proof}
For a nonzero direction $p$ with $p_1\ge0$, choose
\begin{equation}
 \begin{gathered}
 M=\max\{p_1/\rho_1^3,\ |p_2|/\rho_2^2,\ |p_3|/\rho_3\}>0,\\
 \eta=p/M,\qquad h_1=(\eta_1/\rho_1^3)^{1/3},\\
 h_2=(|\eta_2|/\rho_2^2)^{1/2},\qquad h_3=|\eta_3|/\rho_3.
 \end{gathered}
 \label{cert:inverse}
\end{equation}
Each ratio is in $[0,1]$ and at least one equals one, so $\eta=p/M$ lies on the surface of the box, $h\in[0,1]^3$ and $\max_jh_j=1$. Choose $s_2,s_3$ as the signs of $\eta_2,\eta_3$, arbitrary where these vanish. Substitution in \eqref{cert:patches} recovers $\eta$, and \eqref{cert:gradient-general} gives
\begin{equation}
 \frac{\eta}{S(\eta)^{1/4}}=\frac{p/M}{S(p)^{1/4}/M}
 =\frac{p}{S(p)^{1/4}}.
 \label{cert:recovery}
\end{equation}
For $p=\chi$ on the half-boundary this is $\chi$ itself. Conversely, every patch point normalizes to that half-boundary by Lemma~\ref{cert:normalization}. Directions with zero coordinates or tied ratios lie on patch edges, possibly in several patches, which is harmless.
\end{proof}

The next proposition is step~(iii) and states what the computation must deliver.
\begin{prop}[Reduction to a finite verification target]
\label{cert:reduction}
Fix a design with its gains, $\Delta_{\mathrm{cert}}$, $\rho_j$ and matrices passing the exact SPD test, and let $\epsilon>0$ and $m_J>0$. Suppose that for each patch $(j,s_2,s_3)$ of \eqref{cert:patches}, every parameter point $h$ of its closed square, and $\eta=\eta(h;s_2,s_3)$,
\begin{equation}
 \begin{gathered}
 \frac{4\,g(\eta,1,\delta)}{S_\eta}\le-\epsilon\quad\text{for }\delta\in\{-\Delta_{\mathrm{cert}},\Delta_{\mathrm{cert}}\},
 \\
 \frac{4\,H(\eta)}{S_\eta}\ge m_J,
 \end{gathered}
 \label{cert:target}
\end{equation}
with $g$, $H$ from \eqref{cert:gH}. Then \eqref{eq:certificate} holds with these $\epsilon$ and $m_J$.
\end{prop}

\begin{proof}
By Lemmas~\ref{cert:coverage} and~\ref{cert:normalization}, \eqref{cert:target} gives both inequalities of \eqref{eq:certificate} on the half-boundary $\{\chi:S(\chi)=1,\ \chi_1\ge0\}$ for $\sigma=1$ and $\delta=\pm\Delta_{\mathrm{cert}}$. Lemma~\ref{supp:scaling-identities} with $t=-1$ and the oddness of the signed powers give
\begin{equation}
 \begin{gathered}
 S(-\chi)=S(\chi),\quad J(-\chi)=J(\chi),\\
 f(-\chi,-\sigma,-\delta)=-f(\chi,\sigma,\delta),\\
 \nabla S(-\chi)^\top f(-\chi,-\sigma,-\delta)
 =\nabla S(\chi)^\top f(\chi,\sigma,\delta).
 \end{gathered}
 \label{cert:inversion}
\end{equation}
Now fix $\chi$ with $S(\chi)=1$. Only the third component of $f$ depends on $(\sigma,\delta)$, through $-k_3\sigma+\delta$, so the scalar product in \eqref{eq:certificate} is affine in $(\sigma,\delta)$. The admissible pairs, $\sigma\in\spow{\chi_1}{0}$ and $|\delta|\le\Delta_{\mathrm{cert}}$, form a segment or a rectangle with vertices in $\{-1,1\}\times\{-\Delta_{\mathrm{cert}},\Delta_{\mathrm{cert}}\}$, so it suffices to check these vertices. A vertex with $\sigma=1$ requires $\chi_1\ge0$ and is covered directly. A vertex with $\sigma=-1$ requires $\chi_1\le0$; then $-\chi$ lies in the half-boundary, and by \eqref{cert:inversion} the value at $(\chi,-1,\delta)$ equals the value at $(-\chi,1,-\delta)$, which is covered. Likewise, $J(\chi)\ge m_J$ holds directly if $\chi_1\ge0$ and by $J(\chi)=J(-\chi)$ otherwise.
\end{proof}

\subsection{Rational interval evaluation}
\label{interval:evaluation}

Step~(iv) bounds $g$, $H$ and $S_\eta$ on a whole closed parameter rectangle at once, in exact rational arithmetic.

An interval is a closed set $\mathbf x=[x^-,x^+]\subset\R$ with rational endpoints; a rational number is identified with its degenerate interval. An interval \emph{encloses} a function on a set of parameters if it contains the value of the function at every point of the set. The verifier uses interval addition, negation, subtraction and multiplication and the interval square $\operatorname{sq}$ and cube $\operatorname{cb}$, with the standard endpoint formulas of interval arithmetic~\citep{Moore2009}; for instance, the product of $\mathbf x$ and $\mathbf y=[y^-,y^+]$ has as endpoints the smallest and the largest of the four products $x^\pm y^\pm$. Each operation encloses every pointwise result, and $\operatorname{sq}(\mathbf x)$ has lower endpoint zero whenever $0\in\mathbf x$.

The powers $S_\eta^{1/6}$, $S_\eta^{1/8}$ and $S_\eta^{1/4}$ need outward root enclosures, whose endpoints are rounded away from the enclosed roots ($H$ needs no roots). For integers $z\ge0$ and $n\ge1$, let $\operatorname{iroot}(z,n)$ be the integer $w$ satisfying $w^n\le z<(w+1)^n$. For $x^-\ge0$ define
\begin{equation}
 \begin{gathered}
 \operatorname{root}_n(\mathbf x)=
 \Bigl[\frac{\operatorname{iroot}(z^-,n)}{2^{40}},\
 \frac{\operatorname{iroot}(z^+,n)+1}{2^{40}}\Bigr],
 \\
 z^-=\lfloor x^-2^{40n}\rfloor,\quad z^+=\lfloor x^+2^{40n}\rfloor .
 \end{gathered}
 \label{interval:root}
\end{equation}
To verify inclusion, let $w^\pm=\operatorname{iroot}(z^\pm,n)$ and $x\in\mathbf x$. Then
$(w^-)^n\le z^-\le x^-2^{40n}\le x\,2^{40n}$, while
$(w^++1)^n\ge z^++1>x^+2^{40n}\ge x\,2^{40n}$.
The second chain uses the integer gap above $z^+$ and the defining floor inequality. Taking nonnegative $n$th roots and dividing by $2^{40}$ proves that the lower endpoint is no larger than $x^{1/n}$ and the upper endpoint is strictly larger.

Evaluating an expression operation by operation, with each parameter replaced by its interval, therefore yields an interval that encloses the expression on the whole rectangle. This follows by induction over the operations, since every parameter lies in its interval and each operation encloses its result whenever its operands lie in theirs. A parameter that occurs several times can make this interval wider than the range of the expression, never narrower, and intersecting it with a separately proved range keeps the enclosure. The only such range used below is $q_i(\eta)\ge0$, which the SPD check gives before every rectangle evaluation.

In the rest of this subsection, $\Delta=\Delta_{\mathrm{cert}}$. Fix a patch $(j,s_2,s_3)$ and a closed rectangle
$\mathbf B=\mathbf h_{j_1}\times\mathbf h_{j_2}\subseteq[0,1]^2$ in its two
free coordinates $j_1<j_2$, $\{j_1,j_2\}=\{1,2,3\}\setminus\{j\}$. With
$\mathbf h_j=[1,1]$ and the sides of $\mathbf B$ as the other two
$\mathbf h_i$, $\operatorname{eval}(\mathbf B)$ evaluates \eqref{cert:patches}, \eqref{cert:raydata} and \eqref{cert:gH}, with the substitution \eqref{cert:root-removal}, in interval arithmetic:
\begin{align}
 \boldsymbol\eta_1&=\rho_1^3\operatorname{cb}(\mathbf h_1),\quad
 \boldsymbol\eta_2=s_2\rho_2^2\operatorname{sq}(\mathbf h_2),\nonumber\\
 \boldsymbol\eta_3&=s_3\rho_3\mathbf h_3;
 \label{interval:eta}\\
 \mathbf p^{(i)}&=P_i\boldsymbol\eta,\quad
 \mathbf q_i=\sum_{l=1}^3\boldsymbol\eta_l\mathbf p^{(i)}_l,\nonumber\\
 \mathbf q_i&\leftarrow[\max(0,q_i^-),\ q_i^+],\qquad i=1,2,3;
 \label{interval:q}\\
 \mathbf S&=\sum_{i=1}^3\operatorname{sq}(\mathbf q_i),\quad
 \mathbf N=\sum_{i=1}^3\mathbf q_i\,\mathbf p^{(i)},\nonumber\\
 \mathbf H&=3\mathbf N_1\boldsymbol\eta_1+2\mathbf N_2\boldsymbol\eta_2
 +\mathbf N_3\boldsymbol\eta_3;
 \label{interval:SNH}\\
 \mathbf g_\pm&=\mathbf N_1\boldsymbol\eta_2
 +\mathbf N_2\bigl(-k_1\rho_1\mathbf h_1\operatorname{root}_6(\mathbf S)\nonumber\\
 &\quad-k_2s_2\rho_2\mathbf h_2\operatorname{root}_8(\mathbf S)+\boldsymbol\eta_3\bigr)\nonumber\\
 &\quad+\mathbf N_3(-k_3\pm\Delta)\operatorname{root}_4(\mathbf S).
 \label{interval:g}
\end{align}
Matrix--vector products and sums are evaluated componentwise. If $S^-\le0$, which happens exactly when all three clipped intervals $\mathbf q_i$ contain zero, evaluation stops with result \textsc{unknown}; otherwise it returns
\begin{equation}
 g_{\rm up}=\max(g_+^+,\ g_-^+),\quad
 H_{\rm lo}=H^-,\quad
 S_{\rm up}=S^+.
 \label{interval:triple}
\end{equation}

By the induction above, whenever evaluation returns \eqref{interval:triple}, every $h\in\mathbf B$ satisfies
\begin{equation}
 \begin{gathered}
 g(\eta(h),1,\pm\Delta)\le g_{\rm up},\qquad
 H(\eta(h))\ge H_{\rm lo},\\
 0<S_\eta\le S_{\rm up}.
 \end{gathered}
 \label{interval:evaluate}
\end{equation}
Here $S_\eta>0$ also follows from SPD, since $h_j=1$ makes $\eta\ne0$. For positive rational thresholds $\epsilon$ and $m_J$, fixed before the run, a rectangle is accepted precisely when evaluation returns a triple with $4g_{\rm up}/S_{\rm up}\le-\epsilon$ and $4H_{\rm lo}/S_{\rm up}\ge m_J$, compared exactly.

\subsection{Completed cover and uniform margins}
\label{interval:cover}

Step~(v) refines the rectangles until the cover is complete. Initially the pending collection holds the twelve closed unit squares, one per patch, and the accepted collection is empty. At each step the verifier removes one pending rectangle and evaluates it. An accepted rectangle moves to the accepted collection. Any other rectangle, an \textsc{unknown} result included, is bisected on its longer free side, ties resolved in favor of the first free coordinate, and its two closed halves, which share the dividing segment through the exact rational midpoint of that side, return to the pending collection. A rectangle that fails acceptance need not contain a violation, since its enclosures can exceed the true ranges, and halving it typically reduces this excess; no termination theorem for arbitrary data or thresholds is claimed. A run that empties the pending collection within the budget of $200\,000$ evaluations has status \textsc{accepted}; otherwise it is \textsc{incomplete} and inconclusive.

Since every iteration either moves a rectangle unchanged to the accepted collection or replaces it by two closed halves with the same union, the accepted and pending rectangles of each patch cover $[0,1]^2$ at every iteration. After a run with status \textsc{accepted}, every $h$ in every patch square therefore lies in an accepted rectangle of that patch, and \eqref{interval:evaluate} holds at $h$ with the bounds obtained when that rectangle was accepted.

\begin{prop}[Uniform margins from a completed cover]
\label{interval:uniform}
If the run for a design with thresholds $\epsilon,m_J>0$ ends with status \textup{\textsc{accepted}}, then \eqref{cert:target} holds with these $\epsilon$ and $m_J$, and so does \eqref{eq:certificate} by Proposition~\ref{cert:reduction}.
\end{prop}

\begin{proof}
For any patch point $h$ and either endpoint, the completed cover supplies an accepted rectangle containing $h$. Write $g,H,S$ for the true values at $h$, so that $g\le g_{\rm up}$, $H\ge H_{\rm lo}$ and $0<S\le S_{\rm up}$ by \eqref{interval:evaluate}. As $S_{\rm up}>0$, acceptance is equivalent to $g_{\rm up}+\frac\epsilon4S_{\rm up}\le0$ and $H_{\rm lo}-\frac{m_J}4S_{\rm up}\ge0$, and therefore
\begin{equation}
 \begin{gathered}
 g+\frac\epsilon4S\le g_{\rm up}+\frac\epsilon4S_{\rm up}\le0,
 \\
 H-\frac{m_J}4S\ge H_{\rm lo}-\frac{m_J}4S_{\rm up}\ge0.
 \end{gathered}
 \label{interval:leafbounds}
\end{equation}
Dividing by $S/4>0$ gives \eqref{cert:target} at every patch point; Proposition~\ref{cert:reduction} then gives the full boundary statement.
\end{proof}

\begin{proof}[Proof of Proposition~\ref{prop:certificate}]
Both runs used the thresholds $\epsilon=1/100$ and $m_J=1/2$, passed the exact SPD checks and completed with zero pending rectangles. Design A took $90\,406$ rectangle evaluations and accepted $45\,209$ rectangles with at most $22$ bisections each; design B took $92\,566$ evaluations and accepted $46\,289$ rectangles with at most $24$ bisections each. Proposition~\ref{interval:uniform} therefore proves \eqref{eq:certificate} with these margins for both designs.
\end{proof}

The verifier \texttt{verify\_quartic.py}, the matrices of Table~\ref{tab:certificate-matrices} as exact rational files and the run records are available at \url{https://github.com/MartinZapf/Implicit-PID-like}. The verifier needs only the Python standard library, whose \texttt{Fraction} type makes the arithmetic exact and platform independent. Its SHA-256 hash is \texttt{e15aa44f}\allowbreak\texttt{cee9c98c}\allowbreak\texttt{9c343f09}\allowbreak\texttt{fa002459}\allowbreak\texttt{b777ce04}\allowbreak\texttt{cef9e603}\allowbreak\texttt{0c1903fa}\allowbreak\texttt{4bacf381}; each run takes under ten minutes with Python~3.11 on one core.

\section{Omitted proofs}
\label{supp:proofs}
Each subsection names the statement or the sections of the main text that it supports.

\subsection{Lower bound (Proposition~\ref{prop:order-lower-bound})}
\label{supp:lower-bound}

\begin{proof}[Single disturbance in the proof of Proposition~\ref{prop:order-lower-bound}]
With the notation of that proof, a hidden component is half the difference of the two branches on a hidden phase. Write its piece on $[t_j,t_{j+1}]$ as $\Delta T\,\hat p(s)$ with $s=(t-t_j)/T\in[0,1]$. The piece has slopes in $[-\Delta,\Delta]$ if $|\hat p'|\le1$, and zero average and zero convolution with $T-\tau$ if $\int_0^1\hat p\dd s=\int_0^1(1-s)\hat p\dd s=0$. Both hold for $\hat\psi(s)=|s-\frac12|-\frac14$, the shape of $\psi$, and for the cubic $\hat\psi_0(s)=(3s-12s^2+10s^3)/9$, which has $\hat\psi_0(0)=0$, $\hat\psi_0(1)=\frac19$ and the derivative $(1-8s+10s^2)/3\in[-\frac15,1]$ on $[0,1]$. For $0\le\vartheta\le\frac14$, the convex combination $\hat\psi_\vartheta=(1-4\vartheta)\hat\psi_0+4\vartheta\hat\psi$ therefore has both properties, and
\[
 \hat\psi_\vartheta(0)=\vartheta,\qquad \tfrac14-\hat\psi_\vartheta(1)=\tfrac59\big(\tfrac14-\vartheta\big).
\]
Chaining $n$ such pieces from $\vartheta_0=0$, with $\vartheta_{i+1}=\hat\psi_{\vartheta_i}(1)$, gives a continuous hidden component of $n$ pieces that starts at zero and ends at $\Delta T\vartheta_n$, where $\vartheta_n=\frac14(1-(\frac59)^n)$.

The disturbance is built in cycles $m=1,2,\dots$ from any $\phi(0)$. Cycle $m$ begins at a sample $t_{K_m}$, with $K_1=0$ and the level $d_m=\phi(t_{K_m})$, and ends at $t_{K+1}$, where $K=K_m+m$. Its branches are $\phi_\pm=d_m\pm\psi_m$ on the hidden phase $[t_{K_m},t_K]$, with $\psi_m$ the hidden component of $m$ pieces, and $\phi_\pm(t)=d_m\pm(\Delta T\vartheta_m+\Delta(t-t_K))$ on the reveal $[t_K,t_{K+1}]$. The two branches coincide before $t_{K_m}$ and differ by $2\psi_m$ up to $t_K$, so they give the same samples up to index $K$ and the same input $u_K$, whereas \eqref{eq:zoh-x1} gives
\begin{align*}
 x^+_{1,K+1}-x^-_{1,K+1}&=2\Delta T^3\int_0^1(1-s)(\vartheta_m+s)\dd s\\
 &=\Delta T^3\big(\vartheta_m+\tfrac13\big).
\end{align*}
The branch with the larger $|x_{1,K+1}|$ is kept, which fixes the sign of $\psi_m$. It gives $|x_{1,K+1}|\ge\Delta T^3(\frac7{24}-\frac18(\frac59)^m)$, and cycle $m+1$ begins at $K_{m+1}=K+1$.

Each choice depends only on the fixed controller, the initial state and $\phi$ on $[0,t_{K_m}]$, so $\phi$ is defined forward on $[0,\infty)$. It is continuous with slopes in $[-\Delta,\Delta]$, and the closed loop driven by it reproduces the samples of the construction. Hence $\limsup_{k\to\infty}|x_{1,k}|\ge7\Delta T^3/24$. For $x_2$, \eqref{eq:zoh-x2} gives the difference $2\Delta T^2\int_0^1(\vartheta_m+s)\dd s=\Delta T^2(2\vartheta_m+1)$, and keeping the branch with the larger $|x_{2,K+1}|$ gives $\limsup_{k\to\infty}|x_{2,k}|\ge3\Delta T^2/4$. The levels $d_m$ may drift without bound, which \eqref{eq:disturbance-class} allows, since it bounds only the slope.
\end{proof}

\emph{Each bound of Proposition~\ref{prop:order-lower-bound} alone is achieved.} By \eqref{eq:zoh-x1}--\eqref{eq:zoh-x2}, a causal controller can compute, for $k\ge1$, the average $\phibar_{k-1}$ and the weighted average $\tilde\phi_{k-1}=2T^{-2}\int_0^T(T-\tau)\phi(t_{k-1}+\tau)\dd\tau$ from $x_{1,k},x_{2,k},x_{1,k-1},x_{2,k-1}$ and $u_{k-1}$. For $\hat\phi_k=3\phibar_{k-1}-2\tilde\phi_{k-1}$, integration by parts gives
\[
 \phi(t_k)-\hat\phi_k=\int_0^T\Big(\frac{2\tau^2}{T^2}-\frac{\tau}{T}\Big)\dot\phi(t_{k-1}+\tau)\dd\tau,
\]
so $|\phi(t_k)-\hat\phi_k|\le\Delta T\int_0^1|2s^2-s|\dd s=\Delta T/4$.

The law $u_k=-2x_{1,k}/T^2-2x_{2,k}/T-\hat\phi_k$ for $k\ge1$, with any $u_0$, gives $x_{1,k+1}=\int_0^T(T-\tau)(\phi(t_k+\tau)-\hat\phi_k)\dd\tau$, hence $|x_{1,k}|\le(\frac16+\frac18)\Delta T^3=\frac7{24}\Delta T^3$ for $k\ge2$. The law $u_k=-x_{2,k}/T-\hat\phi_k$ gives $x_{2,k+1}=\int_0^T(\phi(t_k+\tau)-\hat\phi_k)\dd\tau$, hence $|x_{2,k}|\le(\frac12+\frac14)\Delta T^2=\frac34\Delta T^2$ for $k\ge2$.

Neither law regulates both states, even for $\phi=0$: the first leaves $x_{2,k}$ alternating in sign, the second leaves $x_{1,k}$ constant.

\subsection{Implicit step (Proposition~\ref{prop:wellposed})}
\label{supp:wellposed}

\begin{proof}[Proof of Proposition~\ref{prop:wellposed}]
On the sliding branch $X=0$,
\eqref{eq:step-first} gives $Y=-a/T$.
The identity $\spow{-a/T}{1/2}=-T^{-1/2}\spow{a}{1/2}$ then makes
\eqref{eq:step-second} equivalent to $\sigma=s$,
admissible if and only if $|s|\le1$.

For $X\ne0$, $\sigma\in\spow{X}{0}$ forces $\sigma=\sgn(X)\in\{-1,1\}$. Fix
either value and eliminate $X=a+TY$ to obtain
\eqref{eq:scalar-residual}. The signed powers
$y\mapsto\spow{a+Ty}{1/3}$ and $y\mapsto\spow{y}{1/2}$ are continuous and
nondecreasing, and the term $y$ makes $R(\cdot\,;\sigma)$ strictly
increasing, with limits $\pm\infty$ as $y\to\pm\infty$. Thus
$R(\cdot\,;\sigma)$ has exactly one root for each $\sigma$; this needs no
differentiability of the signed powers at zero. At the branch threshold,
$R(-a/T;\sigma)=T^2k_3(\sigma-s)$. For $\sigma=1$, consistency requires
$X>0$, that is, $Y>-a/T$. Strict increase of $R(\cdot\,;1)$ makes this
equivalent to $R(-a/T;1)<0$, or $s>1$. Likewise $\sigma=-1$ requires
$Y<-a/T$, equivalent to $s<-1$. Hence exactly one nonzero
branch is admissible when $|s|>1$, and none when $|s|\le1$.
At $s=\pm1$ the corresponding root is $-a/T$, reproducing
the sliding branch.
\end{proof}

\subsection{Planned references and gain scaling}
\label{supp:remarks}

\emph{Deadbeat control and planned references (Sections~\ref{subsec:deadbeat} and~\ref{subsec:small-period}).}
Deadbeat control from $k=0$, $a_{k+1}\equiv0$ in \eqref{eq:applied-input},
has the first input $u_0=-x_{1,0}/T^2-3x_{2,0}/(2T)$, of order $T^{-2}$ if
$x_{1,0}\ne0$. A reference planned over a horizon independent of $T$ and
tracked by \eqref{eq:applied-input} avoids this peak and attains the
conclusions of Theorem~\ref{thm:accuracy} on the plant from an index
that does not depend on the disturbance. For such a reference, let $a_k=r_T(kT)$
for $k\ge0$, where $r_T$ has a locally absolutely continuous derivative,
$r_T(0)=c_0$, $\dot r_T(0)=x_{2,0}$, $r_T=0$ on $[t_H,\infty)$ and
$|\ddot r_T|\le\Gamma$ almost everywhere, with $t_H$ and $\Gamma$ independent of $T$
for all sufficiently small $T$;
let $b_0=x_{2,0}$ and $b_k=(a_k-a_{k-1})/T$ for $k\ge1$. The derivation of
\eqref{eq:physical-identities} uses only \eqref{eq:applied-input}, the
initial values and $b_{k+1}=(a_{k+1}-a_k)/T$, so it holds
for this pair. Inserting it into \eqref{eq:applied-input} gives
\begin{equation*}
 u_k+\phibar_{k-1}=\frac{b_{k+1}-b_k}T-\frac{e_{1,k}}{T^2}-\frac{3e_{2,k}}{2T}.
\end{equation*}
Taylor's theorem gives $|b_{k+1}-b_k|\le\Gamma T$ for $k\ge1$ and
$|b_1-b_0|\le\Gamma T/2$, so the first term is at most $\Gamma$. For $k\ge3$,
\eqref{eq:error-bound} bounds the others by $(C_1+3C_2/2)\Delta T$, and for
$k\le2$ they contain only $g_0$ and $g_1$, bounded as in the proof of
Corollary~\ref{cor:small-period}. For $k\ge\lceil t_H/T\rceil+1$,
$a_k=b_k=0$, and the proof of Theorem~\ref{thm:accuracy} applies with this
index in place of $k_{\mathrm{cap}}$, except for the statement on $v_k$; the
index depends on $t_H$ and $T$ only, not on the disturbance. The proposed
controller is instead a time-invariant dynamic feedback without planning
horizon.

\emph{Gain scaling (Section~\ref{subsec:transformed-recurrence}).} Write $(k_1,k_2,k_3)$ for the gains of a design with $\Lambda=1$ and $f_1$ for their field. The field $f_\Lambda$ of the gains $(\Lambda^{2/3}k_1,\allowbreak\Lambda^{1/2}k_2,\allowbreak\Lambda k_3)$ satisfies $f_\Lambda(\Lambda\zeta,\sigma,\delta)=\Lambda f_1(\zeta,\sigma,\delta/\Lambda)$, with unchanged sign selections, because $\spow{\Lambda\xi}{1/3}=\Lambda^{1/3}\spow{\xi}{1/3}$ and $\spow{\Lambda y}{1/2}=\Lambda^{1/2}\spow{y}{1/2}$. Hence $\zeta\mapsto\zeta/\Lambda$ maps the recurrence of these gains with increments $\delta_k$ onto \eqref{eq:normalized-recurrence} for $\Lambda=1$ with increments $\delta_k/\Lambda$, and Theorem~\ref{thm:capture-recurrence} holds for every $\Lambda>0$ with the slope bound $\Lambda$ times that of $\Lambda=1$ and $V(\zeta/\Lambda)$ in place of $V(\zeta)$. Gains outside the two designs need a new search and verification, either of which may fail.

\subsection{Capture (Section~\ref{sec:capture})}
\label{supp:capture}

\begin{proof}[Proof of Lemma~\ref{cert:geometry}]
Each $\chi^\top P_i\chi$ is a nonnegative convex quadratic and squaring is convex and nondecreasing on $[0,\infty)$, so $S$ is a convex polynomial; its supporting-plane inequality is the first-order condition of convexity. The bound $S(\chi)\ge\sum_i\lambda_{\min}(P_i)^2\|\chi\|^4$ gives coercivity; with continuity and $S(0)=0$ the assertions about $\mathcal K$ follow.
\end{proof}

\begin{proof}[Proof of Lemma~\ref{lem:ilf}]
The radial normalization $\widehat\chi=\chi/S(\chi)^{1/4}$, along straight rays rather than the curves $\gamma\mapsto D(\gamma)\chi$, gives $S(\widehat\chi)=1$ and
\begin{equation}
 J(\chi)=S(\chi)J(\widehat\chi)\ge m_JS(\chi)>0\ \ (\chi\ne0).
 \label{eq:radial-positivity}
\end{equation}
Thus \eqref{cert:radial} gives $\partial_\gamma Q=-J(D(\gamma)^{-1}\zeta)/\gamma<0$. For fixed $\zeta\ne0$, coercivity implies $Q\to+\infty$ as $\gamma\downarrow0$, whereas $Q\to-1$ as $\gamma\to\infty$. Continuity and strict decrease give the unique positive root, and the implicit function theorem gives smoothness off zero. Moreover, $V(\zeta)\le\gamma$ is equivalent to $Q(\zeta,\gamma)\le0$, proving \eqref{eq:sublevels}, also at zero; with $D(\mu)D(\gamma)=D(\mu\gamma)$ it gives $V(D(\mu)\zeta)=\mu V(\zeta)$. Invertible linear images of $\mathcal K$ retain compactness, convexity and an interior neighborhood. The last property gives continuity at zero: given $\gamma>0$, a ball in $D(\gamma/2)\mathcal K$ has $V\le\gamma/2<\gamma$. The bound
$\sup_{D(\gamma)\mathcal K}\|\zeta\|\le\|D(\gamma)\|\max_{\mathcal K}\|\chi\|\to0$ proves shrinkage.
Along the continuous closed loop $\dot\zeta=f(\zeta,\sigma,\delta)$, differentiating \eqref{eq:ilf} gives $\dot V=\nabla S(\chi)^\top f(\chi,\sigma,\delta)/J(\chi)$ at $\chi=D(V)^{-1}\zeta$, at most $-\epsilon/J_{\max}$ by \eqref{eq:certificate}; the origin of the continuous closed loop is therefore asymptotically stable for the certified designs.
\end{proof}

\begin{proof}[Uniform Lyapunov stability in Theorem~\ref{thm:capture-recurrence}]
Given a neighborhood $U$ of zero, choose $\gamma>0$ with $D(\gamma)\mathcal K\subset U$. Continuity of $V$ at zero gives $\eta_0>0$ such that $\|\zeta_{k_0}\|<\eta_0$ implies $V(\zeta_{k_0})<\gamma$, and nonincrease retains every later state in that sublevel set. These choices depend neither on the increment sequence nor on the initial index $k_0\ge2$.
\end{proof}

\emph{Explicit capture bound (Section~\ref{subsec:proof-capture}).} The
proof of Theorems~\ref{thm:capture-recurrence} and~\ref{thm:capture} does
not evaluate $\ell$ and $J_{\max}$. For both designs and every
$\Delta\le\Delta_{\mathrm{cert}}$, $\ell=1/16$ satisfies
$D(\ell)\mathcal K\subset\mathcal C_\Delta$: on $\mathcal K$,
Cauchy--Schwarz gives $\chi^\top(P_1+P_2+P_3)\chi\le\sqrt3<7/4$, which
bounds $|\chi_i|$ by $w_i=(\tfrac74[(P_1+P_2+P_3)^{-1}]_{ii})^{1/2}$, and
$2w_1\ell^3+w_2\ell^2+w_3\ell+k_2(w_1\ell^3)^{1/2}<k_3-\Delta_{\mathrm{cert}}$
holds in exact arithmetic with outward-rounded roots. Further,
$J(\chi)=4\sum_iq_i(\chi)\,\chi^\top H_i\chi$ with
$q_i(\chi)=\chi^\top P_i\chi\ge0$ and $H_i=(P_iW+WP_i)/2$; if $m_iP_i-H_i$
is positive definite, then $J\le4\max_im_i\,S=:\bar JS$, so
$J_{\max}\le\bar J$. Exact Sylvester tests give $(m_1,m_2,m_3)=(19,42,28)$
for design A and $(22,20,30)$ for design B, hence $\bar J=168$ and $120$.
As $\omega$ decreases in $J_{\max}$,
$\underline\omega(\lambda)=\epsilon\lambda/(\bar J(\lambda+\epsilon))\le\omega(\lambda)$,
with $\underline\omega(1/16)=1/19488$ for design A and $1/13920$ for design
B, and $k_{\mathrm{cap}}\le5+\lceil V(\zeta_2)/\underline\omega(1/16)\rceil$;
for $\Lambda\ne1$, $V(\zeta_2/\Lambda)$
replaces $V(\zeta_2)$. The bound is conservative: for design A and
$z_0=(0.15,-0.04,0)$, the initial data of Section~\ref{subsec:sim-orders},
$V(z_0)\approx0.640$ (computed in floating point) gives
$\limsup_{T\downarrow0}t_{k_{\mathrm{cap}}}\lesssim1.25\times10^4$~s,
against a simulated capture time of about $6$~s. The scripts
\texttt{check\_quartic\_consequences.py} and \texttt{bound\_jmax.py} of
the repository repeat the two exact tests.

\begin{rem}[What homogeneity carries over]
\label{rem:homogeneity}
Retained: $f$ is homogeneous of degree $-1$ with respect to the dilation
$D(\gamma)=\operatorname{diag}(\gamma^3,\gamma^2,\gamma)$, and \eqref{eq:normalization}
is the dilation $D(T)^{-1}$, so the implicit Euler step of every length
$T$ becomes the same unit step; one certificate thus covers every $T$
and every $\Lambda$. Not inherited: after
capture, the samples are linear in $\phi$ and independent of the gains
(Section~\ref{sec:accuracy}); their orders $(3,2)$ match those of the
homogeneous continuous design under sampling (as noted after
Theorem~\ref{thm:accuracy}) but
follow from the exact sampled model, not from homogeneity.
\end{rem}

\subsection{Accuracy (Section~\ref{sec:accuracy})}
\label{supp:accuracy}

\emph{The index $k_{\mathrm{acc}}$ cannot be lowered.} For constant $\phi=d$, $g_j=0$ for $j\ge1$, but $e_2=A_Tg_0=(T^2d/2,\,-Td)^\top$, which equals $(x_{1,2},x_{2,2})^\top$ when capture occurs at $k=2$, as it does for $x(0)=0$, $v_0=0$ and $|d|\le Tk_3$. As $\Delta=0$ here, $k_{\mathrm{acc}}=\max\{k_{\mathrm{cap}},3\}$ cannot be replaced by $k_{\mathrm{cap}}$ in Theorem~\ref{thm:accuracy}.

\emph{Between samples in the proof of Theorem~\ref{thm:accuracy}.} Let $h$ have an absolutely continuous derivative
with $|\ddot h|\le M$ almost everywhere on $[t_k,t_{k+1}]$, and let
$E(\tau)=h(t_k+T\tau)-(1-\tau)h(t_k)-\tau h(t_{k+1})$ be its deviation
from the chord. Then
\begin{equation}
 |E(\tau)|\le\frac{MT^2}{2}\tau(1-\tau)\le\frac{MT^2}{8}.
 \label{eq:interpolation-estimate}
\end{equation}
For $k\ge k_{\mathrm{acc}}$, both chord ends obey \eqref{eq:sampled-bounds}.
Since $\ddot x_2=\dot\phi$ on each sampling interval, \eqref{eq:interpolation-estimate}
with $M=\Delta$ bounds $x_2$. The derivative $\dot x_2=u_k+\phi$ has
the average $(x_{2,k+1}-x_{2,k})/T$ over the interval, at most $2C_2\Delta T$ in
magnitude, and differs from it by at most
$T^{-1}\int_{t_k}^{t_{k+1}}|\phi(t)-\phi(s)|\dd s\le\Delta T/2$. Hence
$|u(t)+\phi(t)|\le C_3\Delta T$ on $[t_k,t_{k+1})$, and
\eqref{eq:interpolation-estimate} with $h=x_1$ and $M=C_3\Delta T$
bounds $x_1$.

\begin{proof}[Proof of \eqref{eq:interpolation-estimate}]
The functions $E(\tau)\mp MT^2\tau(1-\tau)/2$ have
absolutely continuous derivatives and, almost everywhere, nonnegative
and nonpositive second derivatives, respectively. Being convex and
concave and zero at $\tau=0$ and $\tau=1$, they are nonpositive and
nonnegative on $[0,1]$, which proves \eqref{eq:interpolation-estimate}.
\end{proof}

\emph{Compensated input.} Inserting
\eqref{eq:physical-identities} into \eqref{eq:applied-input} and using
\eqref{eq:retained-difference} and \eqref{eq:modified-second} gives, for
every $k\ge0$,
\begin{equation}
 \begin{split}
 u_k+\phibar_{k-1}={}&-k_1\spow{a_{k+1}}{1/3}-k_2\spow{b_{k+1}}{1/2}\\
  &+v_{k+1}+\phibar_{k-1}-\frac{e_{1,k}}{T^2}-\frac{3e_{2,k}}{2T},
 \end{split}
 \label{eq:compensated-input}
\end{equation}
the law \eqref{eq:continuous-pid} at
$(a_{k+1},b_{k+1},v_{k+1}+\phibar_{k-1})$ plus a correction that
\eqref{eq:error-bound} bounds by $(C_1+3C_2/2)\Delta T$ for
$k\ge3$.

\begin{proof}[Proof of Corollary~\ref{cor:small-period}]
Every solution of \eqref{eq:generic-step} satisfies $R(Y;\sigma)=0$ in
\eqref{eq:scalar-residual} with $|\sigma|\le1$; as $R(y;\sigma)-y$ is
nondecreasing, $|Y|\le|R(0;\sigma)|\le|b|+T|w|+T^2k_3+Tk_1|a|^{1/3}$.
An update with bounded $(a,b,w)$ thus changes $(a,b,v)$ by $O(T)$, and
\eqref{eq:startup}, $q_2=v_2+\phibar_0$ and
$|\phibar_0-\phi(0)|\le\Delta T/2$ give $(a_2,b_2,q_2)\to z_0$. This
limit, the continuity of $V$ and the bound
$t_{k_{\mathrm{cap}}}<6T+V(a_2,b_2,q_2)/\omega(\ell)$ stated before
Corollary~\ref{cor:small-period} give the bound on the limit superior. For
$k\ge3$, the first four terms of \eqref{eq:compensated-input} are
$-k_1\spow{a_{k+1}}{1/3}-k_2\spow{b_{k+1}}{1/2}+q_{k+1}$. As
$V(a_k,b_k,q_k)=TV(\zeta_k)$ is nonincreasing for $k\ge2$
(Lemma~\ref{lem:decrease}), $(a_{k+1},b_{k+1},q_{k+1})$ lies in
$D(\gamma)\mathcal K$ with $\gamma=V(a_2,b_2,q_2)$ by \eqref{eq:sublevels},
or is zero if $\gamma=0$, so these terms are at most $c_{\mathcal K}\gamma$
in magnitude. For $k\le2$, the $O(T)$ estimate bounds them, and the
correction contains only $g_0$ and $g_1$, whose components are at most
$T^2$ and $2T$ times $\max_{[0,2T]}|\phi|$; it therefore stays bounded.
\end{proof}

\subsection{Simulations (Sections~\ref{sec:simulations} and~\ref{subsec:sim-limits})}
\label{supp:simulations}

\emph{Solver and explicit discretization.} Off the sliding branch,
Brent's method~\citep{Brent1973} solves \eqref{eq:scalar-residual} to a
residual $\rho$ with $|\rho|/T^2\le3.6\times10^{-6}$; the residuals of
two successive steps act on the recurrence of
Section~\ref{sec:capture} like an additional increment of at most
twice this size. The explicit (forward-Euler) discretization of
\eqref{eq:continuous-pid} is
$u_k=-k_1\spow{x_{1,k}}{1/3}-k_2\spow{x_{2,k}}{1/2}+v_k$,
$v_{k+1}=v_k-Tk_3\sgn(x_{1,k})$.

\emph{The cascade.} In the proportional--super-twisting (P-STC) cascade
of \citet{Zapf2025}, the modified implicit super-twisting controller of
\citet{Andritsch2024} acts on $s_k=x_{2,k}+k_px_{1,k}$ with gains $a$
and $b$. With $\mu_k=|s_k|/(T^2b)$ and $\bar\mu_k=\min\{\mu_k,1\}$,
\begin{align*}
 u_k&=-a\psi_k\sgn(s_k)+\nu_{k+1},\\
 \nu_{k+1}&=\nu_k-Tb\bar\mu_k\sgn(s_k),\\
 \psi_k&=\tfrac{Tb}{a}\bar\mu_k-\tfrac{Ta}2
   +T\bigl(\tfrac{a^2}4+b(\mu_k-\bar\mu_k)\bigr)^{1/2}.
\end{align*}
The tuning is $k_p=2$, $a=k_2$ and $b=k_3$, with $\nu_0=0$.

\emph{Runs of Section~\ref{subsec:sim-limits}.} Each run is the load
example over $16$~s with one error. The sampling times are
$T\in\{1,2,5,10,20,50\}$~ms; for the proposed controller, $\beta$ runs
from $0.4$ to $1.25$ in steps of $0.025$; the lags are $0.1$, $0.14$,
$0.17$, $0.25$, $0.5$, $1$ and $2$ times~$T$, with the actuator started
at rest, and the delays $0.1$, $0.14$, $0.16$, $0.25$, $0.5$ and $1$
times~$T$. The earlier implementation and the cascade are run with
the same lags and delays and with $\beta\in\{0.3,0.5,2,3\}$. The
proposed controller counts as on the sliding branch if it
is on it at every sample with $12\le t_k<16$~s, and a run counts as
diverged if $\max\{|x_{1,k}|,|x_{2,k}|\}>10^3$ at some sample or the
solver fails.

%% file: main.bbl
\begin{thebibliography}{24}
\expandafter\ifx\csname natexlab\endcsname\relax\def\natexlab#1{#1}\fi
\providecommand{\url}[1]{\texttt{#1}}
\providecommand{\href}[2]{#2}
\providecommand{\path}[1]{#1}
\providecommand{\DOIprefix}{doi:}
\providecommand{\ArXivprefix}{arXiv:}
\providecommand{\URLprefix}{URL: }
\providecommand{\Pubmedprefix}{pmid:}
\providecommand{\doi}[1]{\href{http://dx.doi.org/#1}{\path{#1}}}
\providecommand{\Pubmed}[1]{\href{pmid:#1}{\path{#1}}}
\providecommand{\bibinfo}[2]{#2}
\ifx\xfnm\relax \def\xfnm[#1]{\unskip,\space#1}\fi
\bibitem[{Acary and Brogliato(2010)}]{AcaryBrogliato2010}
\bibinfo{author}{Acary, V.}, \bibinfo{author}{Brogliato, B.},
  \bibinfo{year}{2010}.
\newblock \bibinfo{title}{Implicit {E}uler numerical scheme and chattering-free
  implementation of sliding mode systems}.
\newblock \bibinfo{journal}{Systems \& Control Letters} \bibinfo{volume}{59},
  \bibinfo{pages}{284--293}.
\newblock \DOIprefix\doi{10.1016/j.sysconle.2010.03.002}.
\bibitem[{Andritsch et~al.(2026a)Andritsch, Koch and
  Moreno}]{AndritschKochMoreno2026}
\bibinfo{author}{Andritsch, B.}, \bibinfo{author}{Koch, S.},
  \bibinfo{author}{Moreno, J.A.}, \bibinfo{year}{2026}a.
\newblock \bibinfo{title}{Model-based disturbance rejection via sliding-mode
  control: {C}ontinuous-time formulation and proper implicit discretization}.
\newblock \bibinfo{howpublished}{arXiv preprint arXiv:2610.02890}.
\newblock \DOIprefix\doi{10.48550/arXiv.2610.02890}.
\bibitem[{Andritsch et~al.(2026b)Andritsch, Koch, Reichhartinger and
  Horn}]{Andritsch2026}
\bibinfo{author}{Andritsch, B.}, \bibinfo{author}{Koch, S.},
  \bibinfo{author}{Reichhartinger, M.}, \bibinfo{author}{Horn, M.},
  \bibinfo{year}{2026}b.
\newblock \bibinfo{title}{A definition of discretization chattering in
  affine-input {SISO} systems}, in: \bibinfo{booktitle}{2026 IEEE 18th
  International Workshop on Variable Structure Systems (VSS)},
  \bibinfo{publisher}{IEEE}. pp. \bibinfo{pages}{155--161}.
\newblock \DOIprefix\doi{10.1109/VSS69650.2026.11655834}.
\bibitem[{Andritsch et~al.(2024a)Andritsch, Watermann, Koch, Reichhartinger,
  Reger and Horn}]{AndritschCDC2024}
\bibinfo{author}{Andritsch, B.}, \bibinfo{author}{Watermann, L.},
  \bibinfo{author}{Koch, S.}, \bibinfo{author}{Reichhartinger, M.},
  \bibinfo{author}{Reger, J.}, \bibinfo{author}{Horn, M.},
  \bibinfo{year}{2024}a.
\newblock \bibinfo{title}{Discretization-chattering-free implementation of
  arbitrary-order sliding mode controllers}, in: \bibinfo{booktitle}{2024 IEEE
  63rd Conference on Decision and Control (CDC)}, \bibinfo{publisher}{IEEE}.
  pp. \bibinfo{pages}{8554--8559}.
\newblock \DOIprefix\doi{10.1109/CDC56724.2024.10886113}.
\bibitem[{Andritsch et~al.(2024b)Andritsch, Watermann, Koch, Reichhartinger,
  Reger and Horn}]{Andritsch2024}
\bibinfo{author}{Andritsch, B.}, \bibinfo{author}{Watermann, L.},
  \bibinfo{author}{Koch, S.}, \bibinfo{author}{Reichhartinger, M.},
  \bibinfo{author}{Reger, J.}, \bibinfo{author}{Horn, M.},
  \bibinfo{year}{2024}b.
\newblock \bibinfo{title}{Modified implicit discretization of the
  super-twisting controller}.
\newblock \bibinfo{journal}{IEEE Transactions on Automatic Control}
  \bibinfo{volume}{69}, \bibinfo{pages}{5620--5626}.
\newblock \DOIprefix\doi{10.1109/TAC.2024.3370494}.
\bibitem[{Barbot et~al.(2020)Barbot, Levant, Livne and Lunz}]{Barbot2020}
\bibinfo{author}{Barbot, J.P.}, \bibinfo{author}{Levant, A.},
  \bibinfo{author}{Livne, M.}, \bibinfo{author}{Lunz, D.},
  \bibinfo{year}{2020}.
\newblock \bibinfo{title}{Discrete differentiators based on sliding modes}.
\newblock \bibinfo{journal}{Automatica} \bibinfo{volume}{112},
  \bibinfo{pages}{108633}.
\newblock \DOIprefix\doi{10.1016/j.automatica.2019.108633}.
\bibitem[{Brent(1973)}]{Brent1973}
\bibinfo{author}{Brent, R.P.}, \bibinfo{year}{1973}.
\newblock \bibinfo{title}{Algorithms for minimization without derivatives}.
\newblock \bibinfo{publisher}{Prentice-Hall}, \bibinfo{address}{Englewood
  Cliffs, NJ}.
\bibitem[{Efimov et~al.(2017)Efimov, Polyakov, Levant and
  Perruquetti}]{Efimov2017}
\bibinfo{author}{Efimov, D.}, \bibinfo{author}{Polyakov, A.},
  \bibinfo{author}{Levant, A.}, \bibinfo{author}{Perruquetti, W.},
  \bibinfo{year}{2017}.
\newblock \bibinfo{title}{Realization and discretization of asymptotically
  stable homogeneous systems}.
\newblock \bibinfo{journal}{IEEE Transactions on Automatic Control}
  \bibinfo{volume}{62}, \bibinfo{pages}{5962--5969}.
\newblock \DOIprefix\doi{10.1109/TAC.2017.2699284}.
\bibitem[{Filippov(1988)}]{Filippov1988}
\bibinfo{author}{Filippov, A.F.}, \bibinfo{year}{1988}.
\newblock \bibinfo{title}{Differential equations with discontinuous righthand
  sides}.
\newblock \bibinfo{publisher}{Kluwer Academic Publishers},
  \bibinfo{address}{Dordrecht}.
\newblock \DOIprefix\doi{10.1007/978-94-015-7793-9}.
\bibitem[{Fridman et~al.(2015)Fridman, Moreno, Bandyopadhyay, Kamal and
  Chalanga}]{Fridman2015}
\bibinfo{author}{Fridman, L.}, \bibinfo{author}{Moreno, J.A.},
  \bibinfo{author}{Bandyopadhyay, B.}, \bibinfo{author}{Kamal, S.},
  \bibinfo{author}{Chalanga, A.}, \bibinfo{year}{2015}.
\newblock \bibinfo{title}{Continuous nested algorithms: The fifth generation of
  sliding mode controllers}, in: \bibinfo{booktitle}{Recent Advances in Sliding
  Modes: From Control to Intelligent Mechatronics}.
  \bibinfo{publisher}{Springer}, \bibinfo{address}{Cham}, pp.
  \bibinfo{pages}{5--35}.
\newblock \DOIprefix\doi{10.1007/978-3-319-18290-2_2}.
\bibitem[{Huber et~al.(2016)Huber, Acary, Brogliato and
  Plestan}]{HuberTwisting2016}
\bibinfo{author}{Huber, O.}, \bibinfo{author}{Acary, V.},
  \bibinfo{author}{Brogliato, B.}, \bibinfo{author}{Plestan, F.},
  \bibinfo{year}{2016}.
\newblock \bibinfo{title}{Implicit discrete-time twisting controller without
  numerical chattering: Analysis and experimental results}.
\newblock \bibinfo{journal}{Control Engineering Practice} \bibinfo{volume}{46},
  \bibinfo{pages}{129--141}.
\newblock \DOIprefix\doi{10.1016/j.conengprac.2015.10.013}.
\bibitem[{Kamal et~al.(2016)Kamal, Moreno, Chalanga, Bandyopadhyay and
  Fridman}]{Kamal2016}
\bibinfo{author}{Kamal, S.}, \bibinfo{author}{Moreno, J.A.},
  \bibinfo{author}{Chalanga, A.}, \bibinfo{author}{Bandyopadhyay, B.},
  \bibinfo{author}{Fridman, L.}, \bibinfo{year}{2016}.
\newblock \bibinfo{title}{Continuous terminal sliding-mode controller}.
\newblock \bibinfo{journal}{Automatica} \bibinfo{volume}{69},
  \bibinfo{pages}{308--314}.
\newblock \DOIprefix\doi{10.1016/j.automatica.2016.02.001}.
\bibitem[{Levant(2005)}]{Levant2005}
\bibinfo{author}{Levant, A.}, \bibinfo{year}{2005}.
\newblock \bibinfo{title}{Homogeneity approach to high-order sliding mode
  design}.
\newblock \bibinfo{journal}{Automatica} \bibinfo{volume}{41},
  \bibinfo{pages}{823--830}.
\newblock \DOIprefix\doi{10.1016/j.automatica.2004.11.029}.
\bibitem[{Levant(2011)}]{Levant2011}
\bibinfo{author}{Levant, A.}, \bibinfo{year}{2011}.
\newblock \bibinfo{title}{Discretization issues of high-order sliding modes}.
\newblock \bibinfo{journal}{IFAC Proceedings Volumes} \bibinfo{volume}{44},
  \bibinfo{pages}{1904--1909}.
\newblock \DOIprefix\doi{10.3182/20110828-6-IT-1002.00489}.
\bibitem[{Livne and Levant(2014)}]{LivneLevant2014}
\bibinfo{author}{Livne, M.}, \bibinfo{author}{Levant, A.},
  \bibinfo{year}{2014}.
\newblock \bibinfo{title}{Proper discretization of homogeneous
  differentiators}.
\newblock \bibinfo{journal}{Automatica} \bibinfo{volume}{50},
  \bibinfo{pages}{2007--2014}.
\bibitem[{Moore et~al.(2009)Moore, Kearfott and Cloud}]{Moore2009}
\bibinfo{author}{Moore, R.E.}, \bibinfo{author}{Kearfott, R.B.},
  \bibinfo{author}{Cloud, M.J.}, \bibinfo{year}{2009}.
\newblock \bibinfo{title}{Introduction to interval analysis}.
\newblock \bibinfo{publisher}{Society for Industrial and Applied Mathematics}.
\newblock \DOIprefix\doi{10.1137/1.9780898717716}.
\bibitem[{P{\'e}rez-Ventura et~al.(2021)P{\'e}rez-Ventura, Mendoza-Avila and
  Fridman}]{PerezVentura2021}
\bibinfo{author}{P{\'e}rez-Ventura, U.}, \bibinfo{author}{Mendoza-Avila, J.},
  \bibinfo{author}{Fridman, L.}, \bibinfo{year}{2021}.
\newblock \bibinfo{title}{Design of a proportional integral derivative-like
  continuous sliding mode controller}.
\newblock \bibinfo{journal}{International Journal of Robust and Nonlinear
  Control} \bibinfo{volume}{31}, \bibinfo{pages}{3439--3454}.
\newblock \DOIprefix\doi{10.1002/rnc.5412}.
\bibitem[{Polyakov et~al.(2015)Polyakov, Efimov and Perruquetti}]{Polyakov2015}
\bibinfo{author}{Polyakov, A.}, \bibinfo{author}{Efimov, D.},
  \bibinfo{author}{Perruquetti, W.}, \bibinfo{year}{2015}.
\newblock \bibinfo{title}{Finite-time and fixed-time stabilization: {I}mplicit
  {L}yapunov function approach}.
\newblock \bibinfo{journal}{Automatica} \bibinfo{volume}{51},
  \bibinfo{pages}{332--340}.
\newblock \DOIprefix\doi{10.1016/j.automatica.2014.10.082}.
\bibitem[{Polyakov et~al.(2023)Polyakov, Efimov and Ping}]{Polyakov2023}
\bibinfo{author}{Polyakov, A.}, \bibinfo{author}{Efimov, D.},
  \bibinfo{author}{Ping, X.}, \bibinfo{year}{2023}.
\newblock \bibinfo{title}{Consistent discretization of homogeneous
  finite/fixed-time controllers for {LTI} systems}.
\newblock \bibinfo{journal}{Automatica} \bibinfo{volume}{155},
  \bibinfo{pages}{111118}.
\newblock \DOIprefix\doi{10.1016/j.automatica.2023.111118}.
\bibitem[{Seeber and Andritsch(2025)}]{SeeberAndritsch2025}
\bibinfo{author}{Seeber, R.}, \bibinfo{author}{Andritsch, B.},
  \bibinfo{year}{2025}.
\newblock \bibinfo{title}{Proper implicit discretization of the super-twisting
  controller---without and with actuator saturation}.
\newblock \bibinfo{journal}{Automatica} \bibinfo{volume}{173},
  \bibinfo{pages}{112027}.
\newblock \DOIprefix\doi{10.1016/j.automatica.2024.112027}.
\bibitem[{Su et~al.(2000)Su, Drakunov and {\"O}zg{\"u}ner}]{Su2000}
\bibinfo{author}{Su, W.C.}, \bibinfo{author}{Drakunov, S.V.},
  \bibinfo{author}{{\"O}zg{\"u}ner, {\"U}.}, \bibinfo{year}{2000}.
\newblock \bibinfo{title}{An {$O(T^2)$} boundary layer in sliding mode for
  sampled-data systems}.
\newblock \bibinfo{journal}{IEEE Transactions on Automatic Control}
  \bibinfo{volume}{45}, \bibinfo{pages}{482--485}.
\newblock \DOIprefix\doi{10.1109/9.847728}.
\bibitem[{Torres-Gonz{\'a}lez et~al.(2017)Torres-Gonz{\'a}lez, Sanchez, Fridman
  and Moreno}]{TorresGonzalez2017}
\bibinfo{author}{Torres-Gonz{\'a}lez, V.}, \bibinfo{author}{Sanchez, T.},
  \bibinfo{author}{Fridman, L.}, \bibinfo{author}{Moreno, J.A.},
  \bibinfo{year}{2017}.
\newblock \bibinfo{title}{Design of continuous twisting algorithm}.
\newblock \bibinfo{journal}{Automatica} \bibinfo{volume}{80},
  \bibinfo{pages}{119--126}.
\newblock \DOIprefix\doi{10.1016/j.automatica.2017.02.035}.
\bibitem[{Xiong et~al.(2019)Xiong, Chen, Jin and Kamal}]{Xiong2019}
\bibinfo{author}{Xiong, X.}, \bibinfo{author}{Chen, W.}, \bibinfo{author}{Jin,
  S.}, \bibinfo{author}{Kamal, S.}, \bibinfo{year}{2019}.
\newblock \bibinfo{title}{Discrete-time implementation of continuous terminal
  algorithm with implicit-{E}uler method}.
\newblock \bibinfo{journal}{IEEE Access} \bibinfo{volume}{7},
  \bibinfo{pages}{175940--175946}.
\newblock \DOIprefix\doi{10.1109/ACCESS.2019.2957282}.
\bibitem[{Zapf et~al.(2025)Zapf, Andritsch, Koch, Fridman and Horn}]{Zapf2025}
\bibinfo{author}{Zapf, M.}, \bibinfo{author}{Andritsch, B.},
  \bibinfo{author}{Koch, S.}, \bibinfo{author}{Fridman, L.},
  \bibinfo{author}{Horn, M.}, \bibinfo{year}{2025}.
\newblock \bibinfo{title}{Implicit discretization of a {PID}-like sliding-mode
  controller for double integrator systems}, in: \bibinfo{booktitle}{2025 IEEE
  64th Conference on Decision and Control (CDC)}, \bibinfo{publisher}{IEEE}.
  pp. \bibinfo{pages}{7939--7944}.
\newblock \DOIprefix\doi{10.1109/CDC57313.2025.11312954}.

\end{thebibliography}
